\documentclass[usenatbib]{mn2e}

\newcommand\nice[1]{#1}    \newcommand\subm[1]{}   % format ``pretty print''
\nice{\newcommand\dosingle[1]{#1}  \newcommand\dodouble[1]{ }} % for single spacing
\subm{\newcommand\dosingle[1]{ }   \newcommand\dodouble[1]{#1}} % for double spacing

\dodouble{\documentclass[usenatbib,referee]{mn2e}}
\usepackage{graphicx}  % graphicx rather than epsf 

\usepackage{amsfonts}
\usepackage{mathpi}

\usepackage{mn_bmath} % monthly notices bold math style

 \usepackage{hyperref}
\nice{ 
\providecommand{\eprint}[1]{\href{http://arxiv.org/abs/#1}{{\tt [arXiv:#1]}}
}

\providecommand{\url}[1]{\href{#1}{#1}}
}

\providecommand{\adsurl}[1]{} 

\def\SSS{Sect.~}
\newcommand\apjs{ApJSupp}                 % {Ap. J.}
\newcommand\aap{A\&A}            % {A. \& A.} 
\newcommand\mnras{MNRAS}
\newcommand\nat{Nature}
\newcommand\cqg{CQG}
\newcommand\pra{Phys.~Rev.~A}
\newcommand\prd{Phys.~Rev.~D}
\newcommand\physrep{Phys. Rep.}
\newcommand\BASI{Bull. Astr. Soc. India}
\newcommand\AmJPhys{Am. J. Phys.}
\newcommand\FoundPhys{Found. Phys.}

\nice{ \newcommand\mycaptionfont{\protect\footnotesize} }

\def\gtapprox{\,\lower.6ex\hbox{$\buildrel >\over \sim$} \, }
\def\ltapprox{\,\lower.6ex\hbox{$\buildrel <\over \sim$} \, }
\def\propapprox{\,\lower.6ex\hbox{$\buildrel \propto\over \sim$} \, }

\def\arcs{\ifmmode {'' }\else $'' $\fi}     %Arc seconds%
\def\arcm{\ifmmode {' }\else $' $\fi}       %Arc minutes%

\def\fr7{7$ \hskip -0.9ex \vrule height0.8ex width0.8ex depth-0.73ex
                                                     \hskip0.1ex$}
\def\frtoday{Le\space\number\day\space\ifcase\month\or
  janvier\or f\'evrier\or mars\or avril\or mai\or juin\or
  juillet\or ao\^ut\or septembre\or octobre\or novembre\or 
d\'ecembre\fi\space \number\year}

\newcommand\hGpc{\mbox{$h^{-1}$ Gpc}}

\newcommand\Omm{\Omega_{\mbox{\rm \small m}}}%%% EDITOR modify as desired 
\newcommand\Omb{\Omega_{\mbox{\rm \small b}}}%%% EDITOR modify as desired 
\newcommand\Omtot{\Omega_{\mbox{\rm \small tot}}}%%% EDITOR modify as desired 

\newcommand\Nother{N_{\mbox{\rm other}}}%%% EDITOR modify as desired 

\begin{document}

\title[Homotopy symmetry in twin paradox]{Homotopy symmetry in the
  multiply connected twin paradox of special relativity}

\author[B. F. Roukema \& S. Bajtlik]{Boudewijn F. Roukema$^1$
and Stanislaw Bajtlik$^2$
%\and Marek Biesiada \inst{3}
%\and Agnieszka Szaniewska \inst{1}
%\and Helena Jurkiewicz \inst{1}
%}
%\address{
\\
$^1$ Toru\'n Centre for Astronomy, Nicolaus Copernicus University,
ul. Gagarina 11, 87-100 Toru\'n, Poland  
\\
%}
%\address{
$^2$ Nicolaus Copernicus Astronomical Center, 
  ul. Bartycka 18, 00-716 Warsaw, Poland
}
%\ead{boud@astro.uni.torun.pl}  %%% no spammable email address on astro-ph!

%\def\today{\frtoday}

\date{\frtoday}
%\date{\today}

%\titlerunning{Homotopy symmetry in twin paradox}
%\authorrunning{Roukema \& Bajtlik}

\maketitle % MNRAS positioning of abstract

\begin{abstract}
%WCWC word count
In a multiply connected space, the two twins of the special relativity twin
paradox move with constant relative speed and meet a second time without
acceleration.  The twins' situations appear to be symmetrical
despite the need for one to be younger due to time dilation. 
Here, the suggestion that the apparent symmetry is
broken by homotopy classes of the twins' worldlines is reexamined using
space-time diagrams. It is found that each twin finds her own spatial path
to have zero winding index and that of the other twin to have unity winding
index, i.e. the twins' worldlines' relative homotopy classes are symmetrical.
Although the twins' apparent symmetry is in fact 
broken by the need for the non-favoured
twin to non-simultaneously identify spatial domain boundaries, the non-favoured
twin {\em cannot} detect her disfavoured state if she only measures 
the homotopy classes
of the two twins' projected worldlines, contrary to what was previously
suggested. We also note that 
for the non-favoured twin, the fundamental domain can be chosen by
identifying time boundaries (with a spatial offset) 
instead of space boundaries (with a temporal offset).
%WCWC word count
\end{abstract}

%\noindent{\it Keywords\/} cosmology of theories beyond the SM, 
%physics of the early universe

\dodouble{\clearpage} %% otherwise first text section is funny

%%%%%%%%%%%%%%%%%%%%%%%%%%%%%%%%%%%%%%%%%%%%%%%%%%%%%%%%%%%%%%%%%%%
%% Figure section. Defined early so that position in output can
%% be easily moved towards earlier pages if LaTeX wants to put
%% them all at the end...
%%%%%%%%%%%%%%%%%%%%%%%%%%%%%%%%%%%%%%%%%%%%%%%%%%%%%%%%%%%%%%%%%%%

\newcommand\fspacetimeone{
\begin{figure}  % [ht]
\centering
\includegraphics[width=8cm]{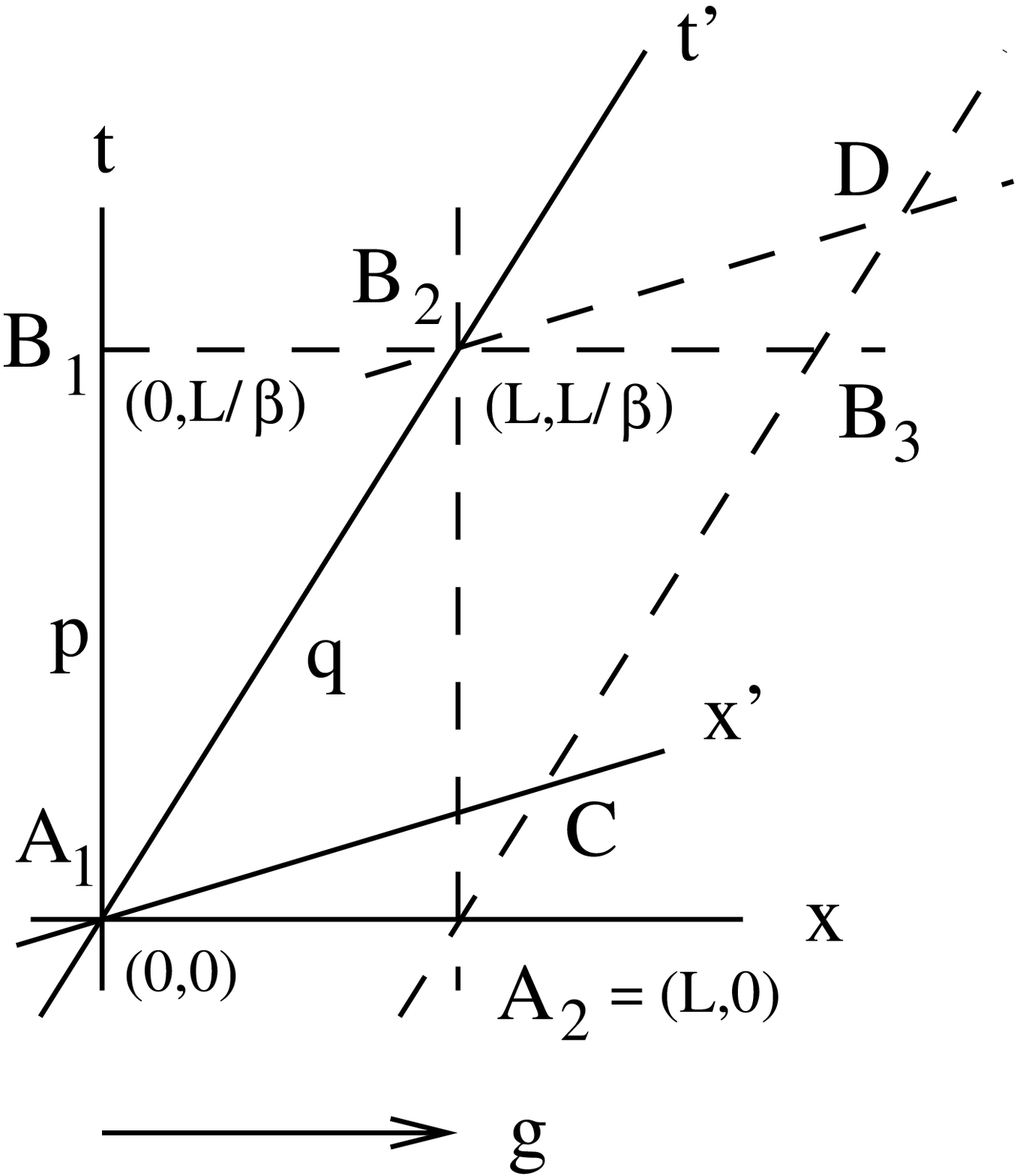}
\caption[stationary twin]{ 
\mycaptionfont {Minkowski covering space space-time for a twin 
(hereafer, the ``leftmoving twin'') with worldline $t$ and 
coordinate system
$(x,t)$, made multiply connected in each spatial section at constant 
time $t$ via the generator (translation) $g$. 
A twin moving to the right at constant relative velocity $\beta$
(hereafter, the ``rightmoving twin'') has
worldline $t'$ and simultaneity axis $x'$ where $t'\equiv 0$, 
defining her coordinate system $(x',t')$.  Space-time points 
A$_1$ and A$_2$ are identical space-time events under the generator $g$, i.e.
the single event may in general be called A.
Similarly, space-time points 
B$_1$, B$_2$ and B$_3$ are a single space-time event B.  Space-time events
C and D are distinct from each other and from A and B; C and D occur at
the same spatial location for the rightmoving twin.
The worldlines of the leftmoving and rightmoving twins are $p$ and $q$ 
respectively.
}
\label{f-spacetimeone}
}
\end{figure}
} % of \def\fspacetimeone

\newcommand\fcylinderone{
\begin{figure}  % [ht]
\centering
\includegraphics[width=35mm]{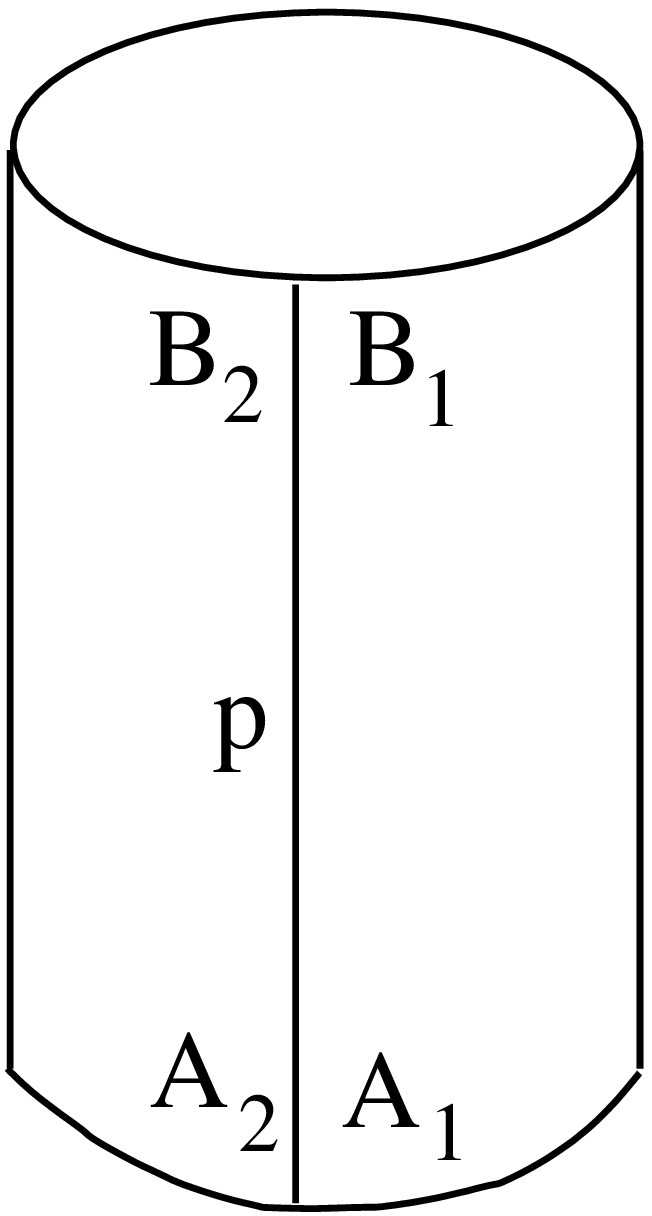}
\caption[stationary twin cylinder]{ 
\mycaptionfont {A space-time region, with constant space and time
boundaries according to the leftmoving twin's view of 
space-time, shown in Figure~\protect\ref{f-spacetimeone},
embedded in 3-D Euclidean space 
and projected onto the page, or informally, 
``rectangle A$_1$A$_2$B$_2$B$_1$ rolled up 
into a cylinder and stuck together to make it multiply connected''.
The leftmoving twin's worldline $p$ is shown.
}
\label{f-cylinderone}
}
\end{figure}
} % of \def\fcylinderone

\newcommand\fspacetimetwo{
\begin{figure}  % [ht]
\centering
\includegraphics[width=8cm]{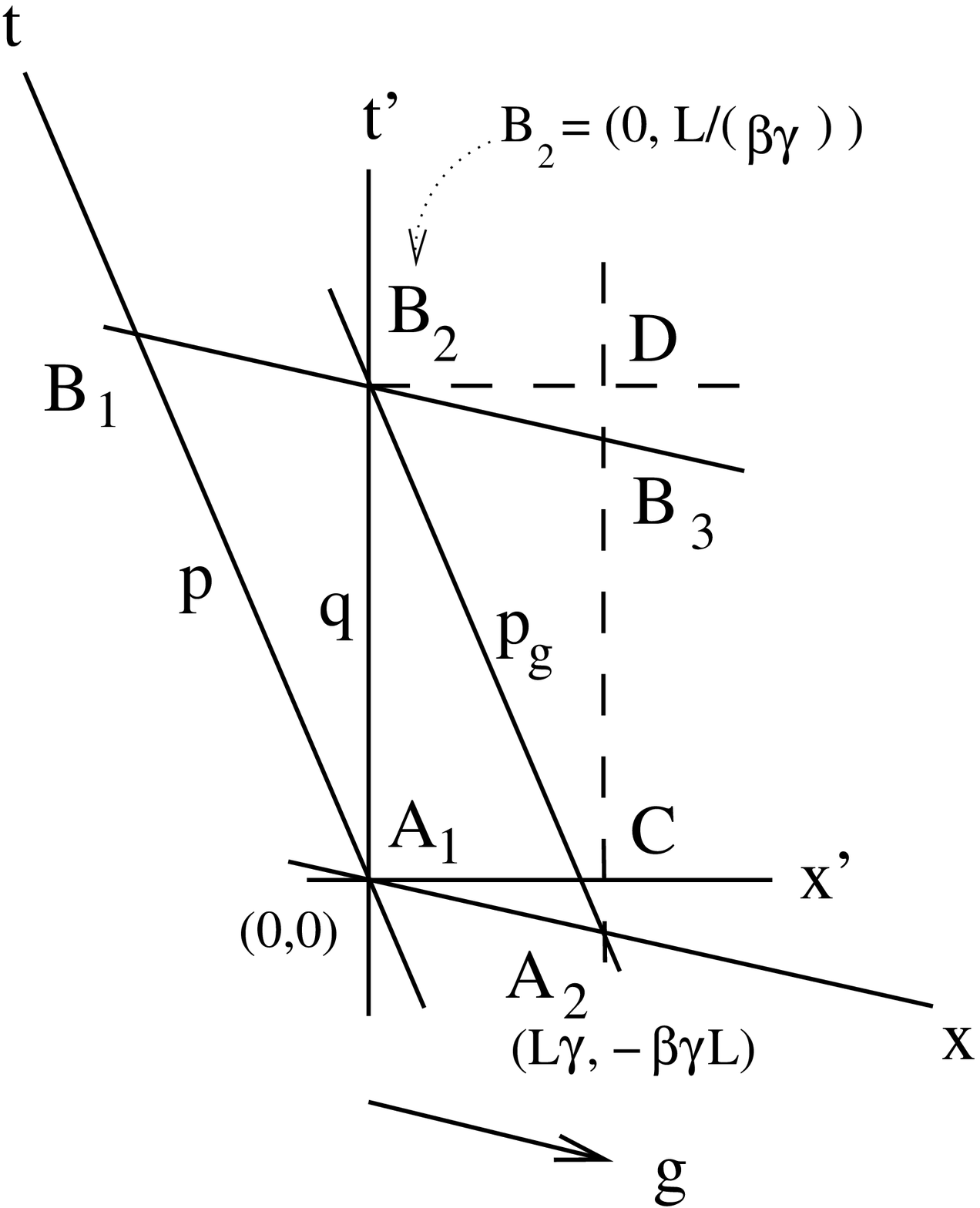}
\caption[moving twin]{ 
\mycaptionfont {Identical space-time 
to Figure~\protect\ref{f-spacetimeone}, but shown in 
the rest frame of the rightmoving twin.
The identities due to the generator $g$ remain correct: A$_1=$A$_2$ and 
B$_1=$B$_2=$B$_3$, {\em even
 though they are non-simultaneous}; $g$ could
be described as a non-simultaneous generator in the rightmoving twin's 
reference frame. The worldline $p$ of the leftmoving twin is mapped by
the generator to a physically equivalent copy, $p_g$, running from A$_2$
to B$_2$.
Figure~\protect\ref{f-cylindertwo} helps show this is possible by using our
three-dimensional intuition.
}
\label{f-spacetimetwo}
}
\end{figure}
} % of \def\fspacetimetwo

\newcommand\fspacetimetwoconst{
\begin{figure}  % [ht]
\centering
\includegraphics[width=8cm]{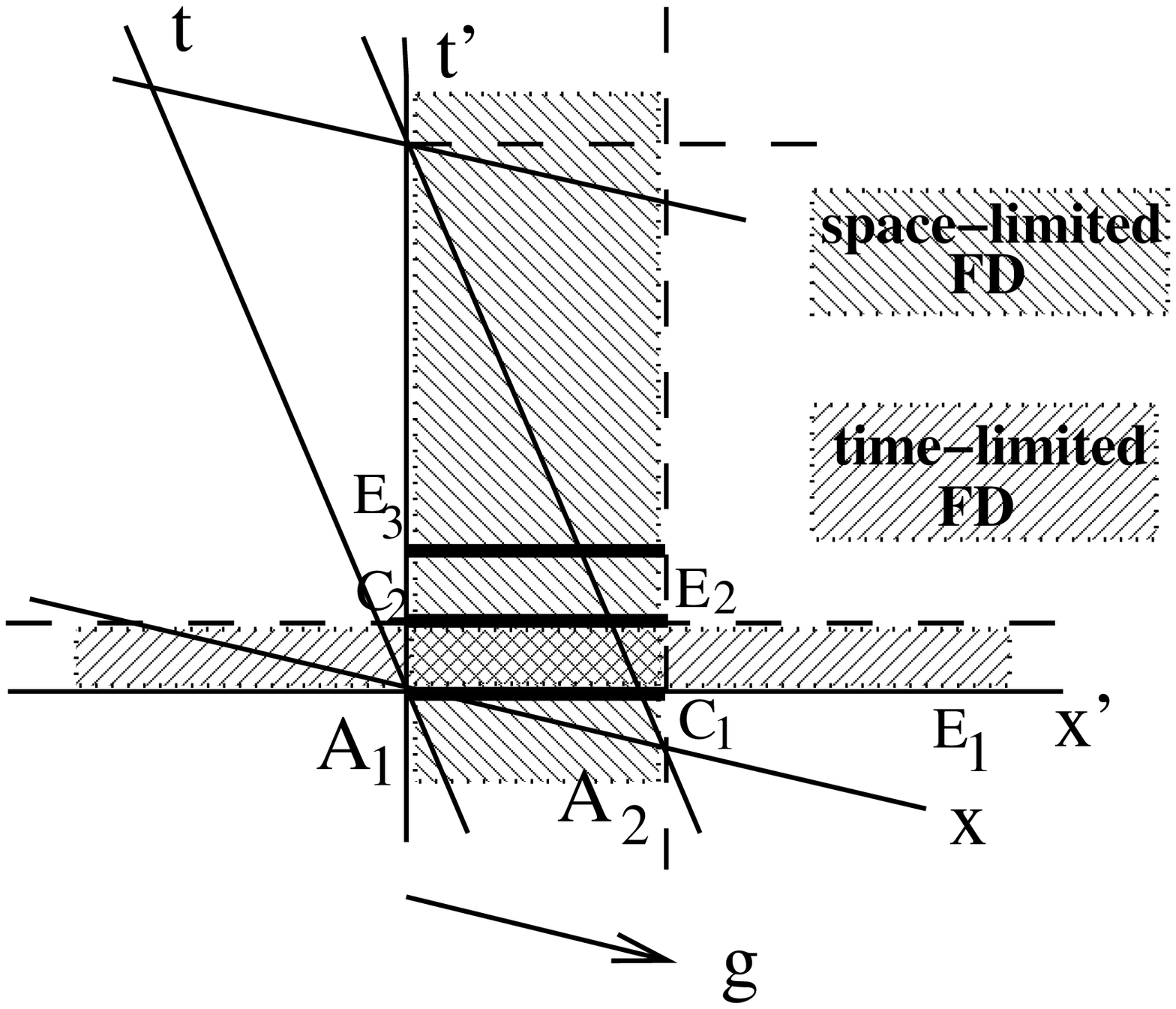}
\caption[moving twin]{ 
\mycaptionfont {Identical space-time 
to Figure~\protect\ref{f-spacetimetwo}, again in the reference frame of
the rightmoving twin, showing 
part of a hypersurface at constant time for the rightmoving twin 
as thick horizontal line segments. Space-time events C$_1$ and C$_2$ are
identical to one another 
(the generator $g$ identifying equal space-time events is illustrated); 
space-time events E$_1$, E$_2$ and E$_3$ are identical to one another. 
A is a single space-time
event, C is a single space-time event, and E is a single space-time event.
The fundamental domain (FD) can be written as a space-limited region with
unconstrained time (shaded with a backslash ``$\backslash$'') and a time offset
when matching boundaries. 
It can also be written as a {\em time}-limited region with
unconstrained space (shaded with a forward slash ``/'') and a space offset
when matching boundaries. Within this FD, consider an observer at rest in
this frame, capable of making metric measurements, 
located at space-time event A and shown
in the covering space at A$_1$. For this observer, 
the events C and E happen ``over there 
and simultaneously to A''. The observer will arrive at those spatially distant
events without any motion, simply by waiting. 
See Fig.~\protect\ref{f-cylindertwo_tperiodic} for a cylindrical representation
of the time-limited FD.
}
\label{f-spacetimetwoconst}
}
\end{figure}
} % of \def\fspacetimetwoconst

\newcommand\fcylindertwo{
\begin{figure}  % [ht]
\centering
\includegraphics[width=25mm]{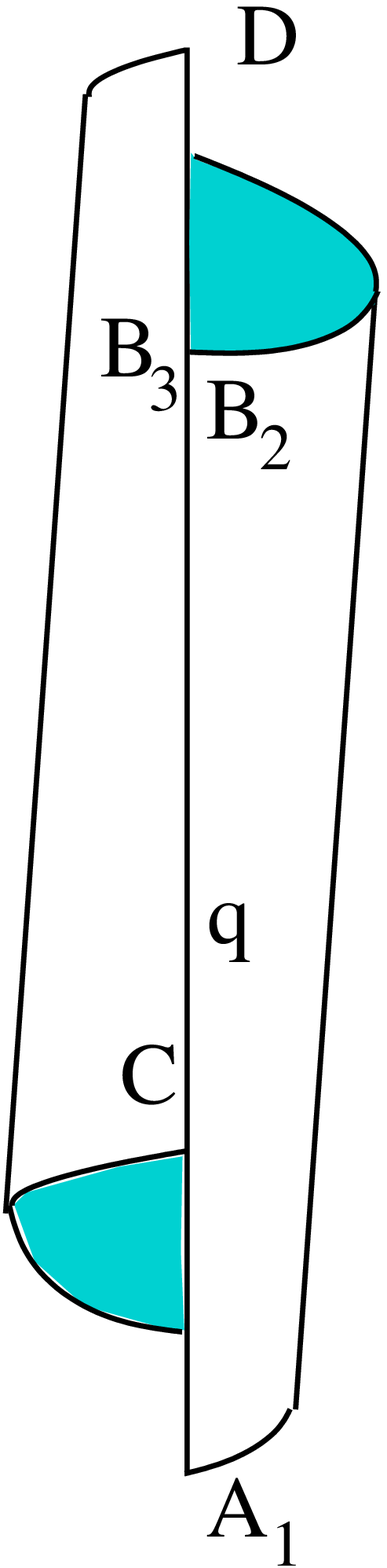}
\caption[moving twin cylinder]{ 
\mycaptionfont {A space-time region
with constant space and time
boundaries according to the rightmoving twin's view of 
space-time, shown in Figure~\protect\ref{f-spacetimetwo},
embedded in 3-D Euclidean space and projected 
onto the page, or informally, ``rectangle A$_1$CDB$_2$ 
rolled up and stuck together to make it multiply connected''. 
{\em The spatial boundaries of this region, 
  $\overline{\mbox{A$_1$B$_2$}}$ and $\overline{\mbox{CD}}$,
  are offset by a time interval A$_2$C $= \beta \gamma L$
  before being matched: the result is not a cylinder.}
Note that in space-time, there are two geodesics joining space-time events 
A$_1$ and $C$: one at constant spatial position 
(appearing vertical here), and
one at constant time (appearing nearly horizontal, but sloped at a moderate
angle in this projection), 
and similarly for B$_2$ and $D$. However, only one of these two 
geodesics --- the vertical (timelike) one ---
can be a worldline of a physical (non-tachyonic) particle,
so there is no causality violation.
A similar diagram to this one has earlier been published in Figure~5b 
in \protect\cite{Wucknitz04}.
The rightmoving twin's worldline $q$ from A$_1$ to B$_2$ is shown.
}
\label{f-cylindertwo}
}
\end{figure}
} % of \def\fcylindertwo

\newcommand\fcylindertwotperiodic{
\begin{figure}  % [ht]
\centering
\includegraphics[width=80mm]{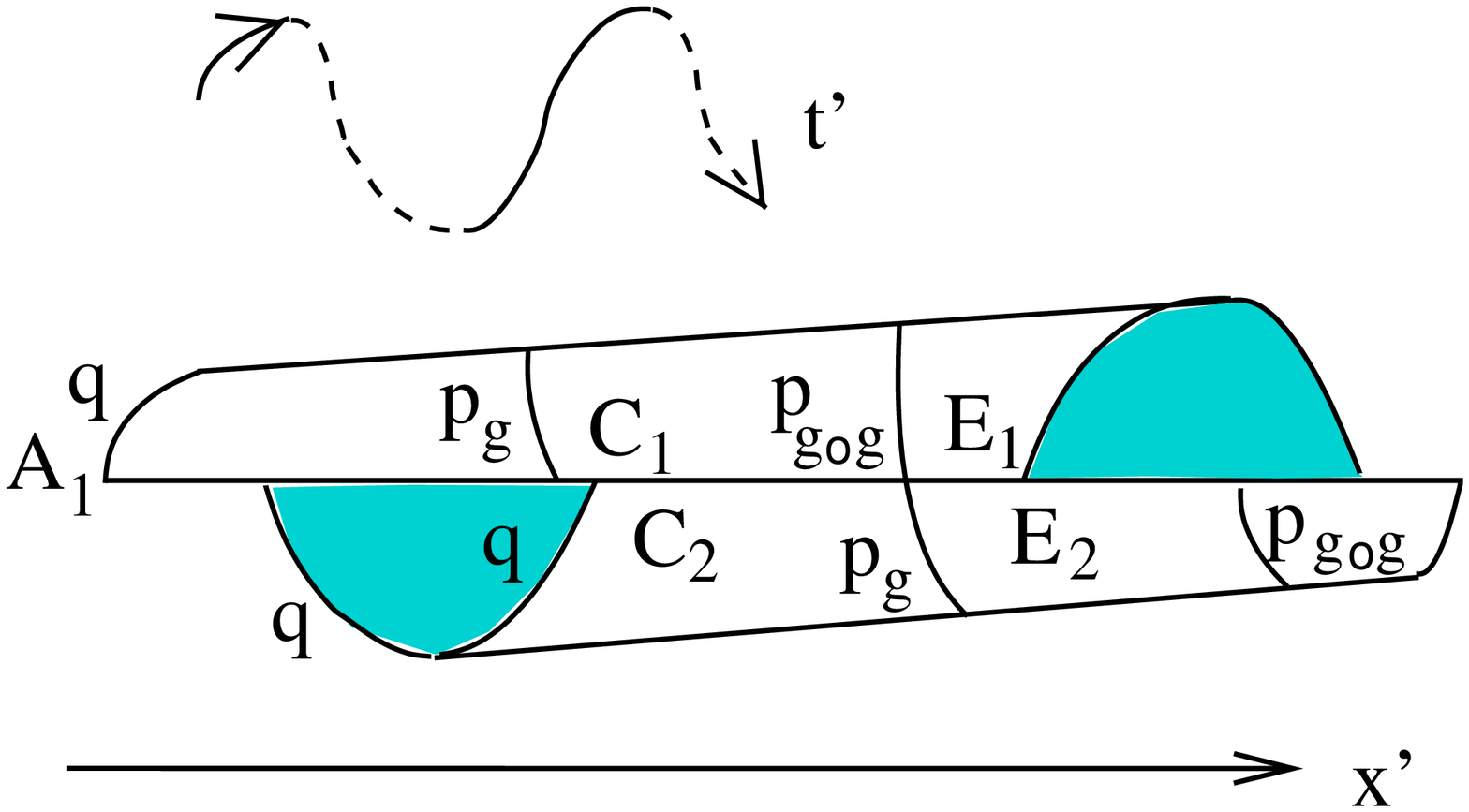}
\caption[moving twin cylinder with time FD]{ \mycaptionfont {A
    time-limited choice of the fundamental domain according to the
    rightmoving twin's view of space-time, embedded in 3-D Euclidean
    space and projected onto the page.  See
    Fig.~\protect\ref{f-spacetimetwoconst} and
    \SSS\protect\ref{s-tperiodic}.  C$_1$ and C$_2$ are a single
    space-time location; E$_1$ and E$_2$ are a single space-time
    location. Time boundaries are identified to one another with a
    spatial offset.  Some of the multiple topological images of the
    worldlines of the leftmoving and rightmoving twins are labelled as
    previously, i.e. $p_g$ and $q$ respectively, as well as $p_{g
      \circ g} \equiv g(p_g)$. The positive spatial and time
    directions are shown as $x'$ and $t'$ respectively.  }
\label{f-cylindertwo_tperiodic}
}
\end{figure}
} % of \def\fcylindertwotperiodic

\newcommand\fcylinderonehomotopy{
\begin{figure}  % [ht]
\centering
\includegraphics[width=60mm]{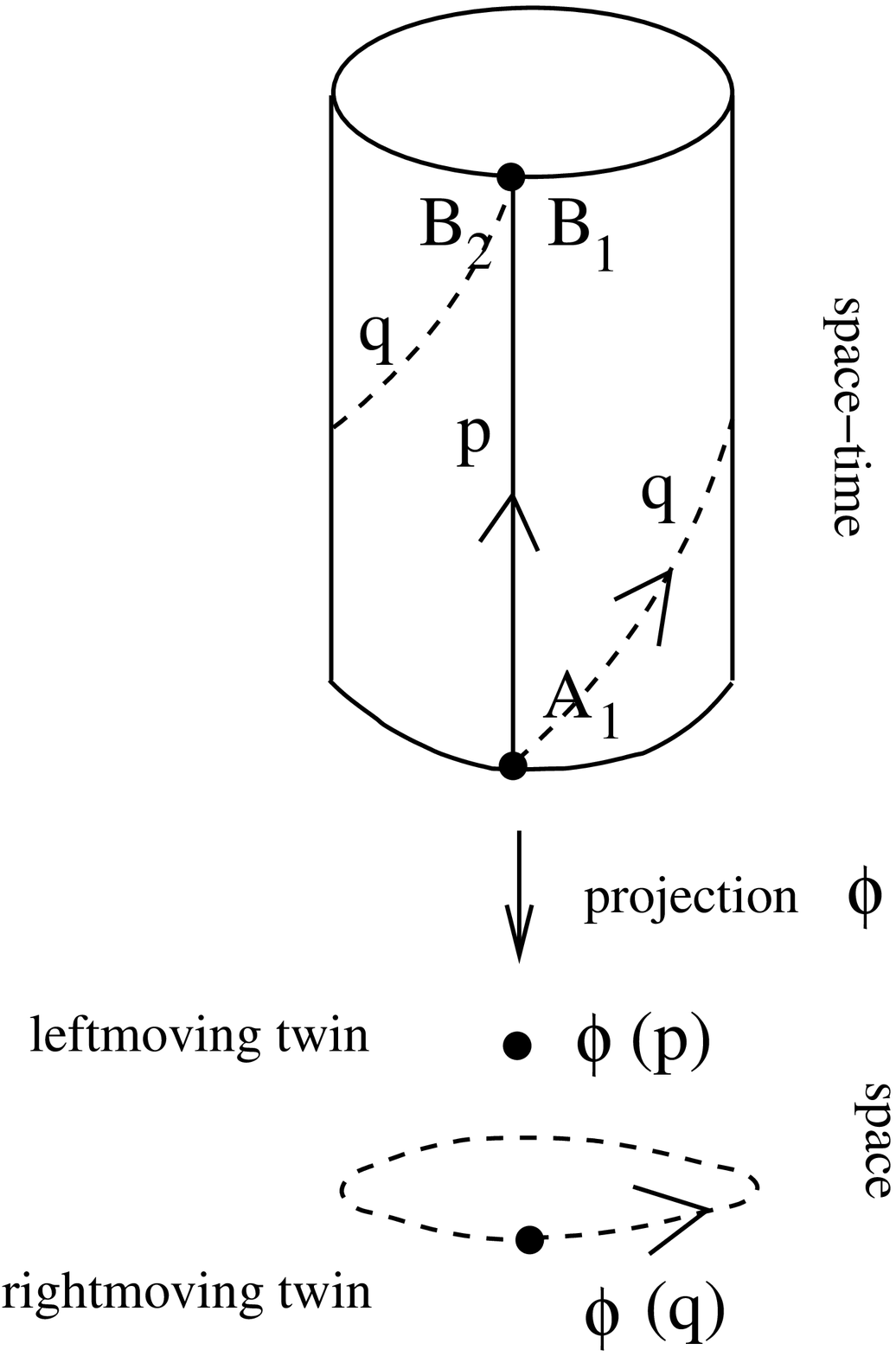}
\caption[stationary twin cylinder homotopy]{ 
\mycaptionfont {Space-time as viewed by the leftmoving twin, 
as in Figure~\protect\ref{f-cylinderone}, 
including the worldlines $p$ and $q$ of the leftmoving and
rightmoving twins respectively, corresponding to
$\overline{\mbox{A$_1$B$_1$}}$ 
and
$\overline{\mbox{A$_1$B$_2$}}$ in the covering space-time as shown in 
Figure~\protect\ref{f-spacetimeone}, 
and showing their projections from space-time into space under the
projection $\phi$ given in Eq.~(\protect\ref{e-phi-L}). Arrows
indicate increasing proper time along each worldline. The worldlines
are given algebraically in \SSS\protect\ref{s-worldlines}.
}
\label{f-cylinderone_homotopy}
}
\end{figure}
} % of \def\fcylinderonehomotopy

\newcommand\fcylindertwohomotopy{
\begin{figure}  % [ht]
\centering
\includegraphics[width=60mm]{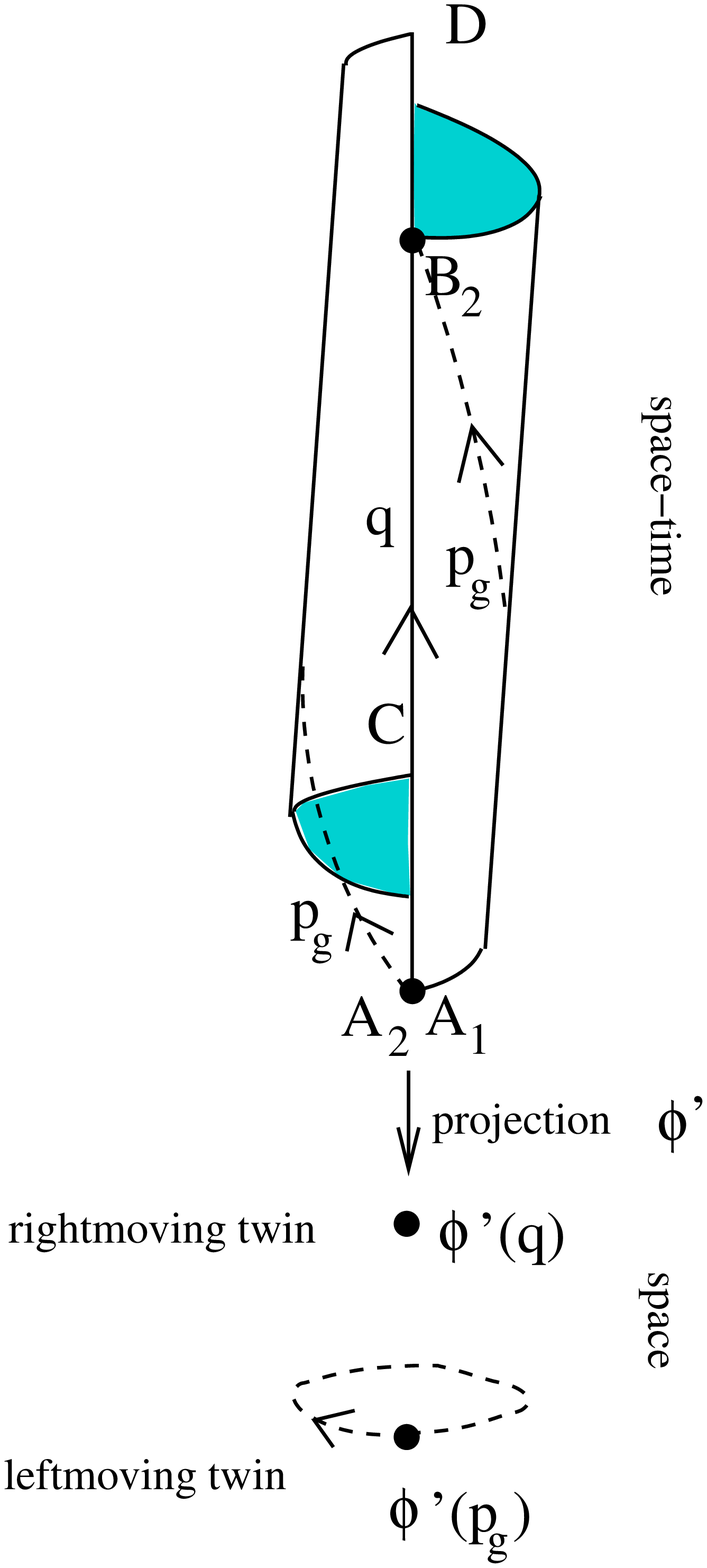}
\caption[stationary twin cylinder homotopy]{ 
\mycaptionfont {Space-time as viewed by the rightmoving twin, 
as in Figure~\protect\ref{f-cylindertwo},
including the worldlines $q, p_g$ of the 
leftmoving and rightmoving twins respectively. These correspond to 
$\overline{\mbox{A$_1$B$_2$}}$ and
$\overline{\mbox{A$_2$B$_2$}}$ in the covering space-time as shown in 
Fig.~\protect\ref{f-spacetimetwo}. 
Also shown are their projections from space-time into 
space under the projection $\phi'$ given in 
Eq.~(\protect\ref{e-phi-R}). 
Arrows indicate increasing proper time along each worldline.
The worldlines
are given algebraically in \SSS\protect\ref{s-worldlines}.
}
\label{f-cylindertwo_homotopy}
}
\end{figure}
} % of \def\fcylindertwohomotopy

\newcommand\fcylinderoneeasy{
\begin{figure}  % [ht]
\centering
\includegraphics[width=45mm]{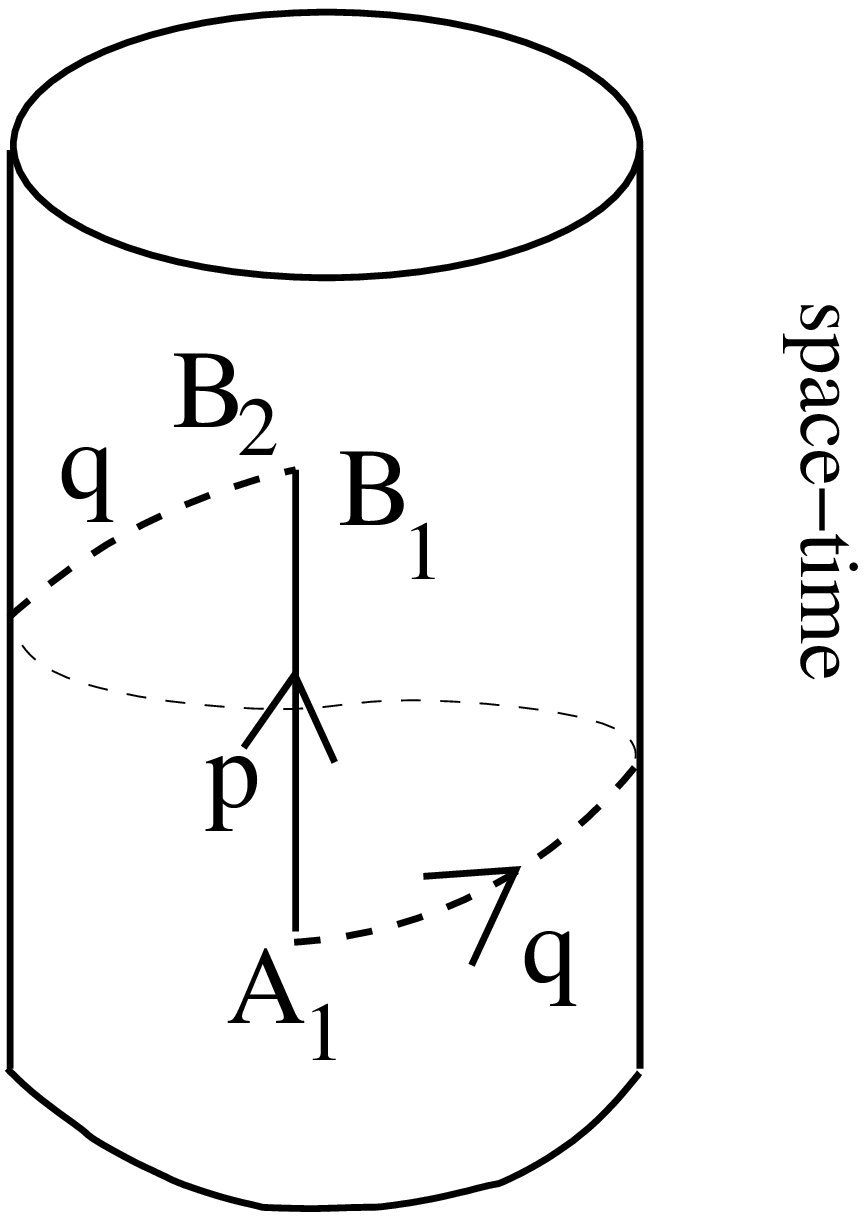}
\caption[stationary twin cylinder simpler]{ 
\mycaptionfont {As for Figure~\protect\ref{f-cylinderone_homotopy},
but showing that the two worldlines together form
a single closed loop in space-time. Neither worldline
constitutes a closed curve alone, prior to projection.
}
\label{f-cylinderone_easy}
}
\end{figure}
} % of \def\fcylinderoneeasy

% \newcommand\fsynchone{
% \begin{figure}  % [ht]
% \centering
% \includegraphics[width=8cm]{synchone}
% \caption[stationary twin cylinder]{ 
% \mycaptionfont {Synchronisation of  leftmoving twin's view of 
% space-time, shown in Fig.~\protect\ref{f-spacetimeone},
% in 3-D Euclidean space and projection onto the page, or informally, ``rolling 
% up rectangle A$_1$A$_2$B$_2$B$_1$ into a cylinder to make it multiply connected''.
% }
% \label{f-synchone}
% }
% \end{figure}
% } % of \def\fsynchone

%\newcommand\fsrexpand{
%\begin{figure}  % [ht]
%\centering
%\includegraphics[width=8cm]{srexpand}
%\caption[moving twin]{ 
%\mycaptionfont {Identical space-time diagram 
%to Fig.~\protect\ref{f-spacetimeone}, except the spatial section 
%for the leftmoving twin is expanding: $g$ is not a constant translation
%with time, and the world line A$_1$B$_1$ matches the world line 
%A$_2$B$_3$ (instead of A$_2$B$_2$, which is for the static case).
%}
%\label{f-srexpand}
%}
%\end{figure}
%} % of \def\fsrexpand

%%%%%%%%%%%%%%%%%%%%%%%%%%%%%%%%%%%%%%%%%%%%%%%%%%%%%%%%%%%%%%%%%%%

\section{Introduction}
Since the confirmation by Wilkinson Microwave Anisotropy Probe 
(WMAP) observations of the cosmic microwave background that 
the large scale ($r > 10 {\hGpc}$ in comoving units) 
auto-correlation function of temperature fluctuations is close to
zero 
[fig.~16, \citealt{WMAPSpergel} (orthogonally projected spatial scale);  
 fig.~1 of \citealt{RBSG08} (non-projected spatial scale)] 
and difficult to reconcile with a standard cosmic concordance
model 
({\SSS}7 of \citealt{WMAPSpergel}; though see also \citealt{EfstNoProb03b})
there has recently been considerable 
observational interest in understanding multiply
connected models of the Universe.
The Poincar\'e dodecahedral space (PDS) has been proposed as a better
model of the 3-manifold of comoving space 
rather than an ``infinite'' flat
space
(e.g. \citealt{LumNat03,RLCMB04,Aurich2005a,Aurich2005b,Gundermann2005,RBSG08};
though see also \citealt{NJ07}).
%(e.g. \citealt{LumNat03}; 
%\citealt{RLCMB04}; \citealt{Aurich2005a};
%\citealt{Aurich2005b}; \citealt{Gundermann2005}; ).
Curvature estimates are consistent with the model: 
analysis of the 3-year WMAP data \citep{WMAPSpergel06} results in 
best estimates of the total density parameter 
$\Omtot = 1.010_{-0.009}^{+0.016}$ (when combined with HST key project on 
$H_0$ data) and
$\Omtot = 1.015_{-0.016}^{+0.020}$ (when combined with Supernova Legacy 
Survey data), 
or together with the third, fourth and fifth acoustic peaks,
as estimated from their
observations near the South Galactic Pole using the
Arcminute Cosmology Bolometer Array Receiver 
(ACBAR), $\Omtot = 1.03^{+0.04}_{-0.06}$ is obtained \citep{ACBAR08},
consistently with that expected from the
PDS analyses, which require positive curvature in this range of 
$\Omtot$ values. 

It has also recently become noticed that global topology in a universe
containing density perturbations can, in principle, have at least some
effect on the metric, even though the effect is expected to be small.
At the present epoch, in the case of a 
$\mathbb{T}^3$ model of fundamental lengths 
which are slightly unequal by a small fraction $\delta$,
a new term appears in the acceleration equation,
causing the scale factor to accelerate or decelerate in the 
directions of the three fundamental lengths
in a way that tends to equalise them \citep{RBBSJ06}.

In this context, the properties and implications 
of the twin paradox of special relativity
in a multiply connected space need to be correctly understood.
It has already been 
shown \citep{BransStewart73,Peters83,Peters86,Low90,UzanTwins02,LevinTwins01}
that resolving the twin paradox of special 
relativity in a multiply-connected ``Minkowski'' space-time 
implies new understanding of the paradox relative to the case in 
simply connected Minkowski space. 

Moreover, it is known that, at least in the case of a static space
with zero Levi-Civita connection, multiple connectedness implies a
favoured space-time splitting. This
should correspond to the comoving reference frame 
\citep{UzanTwins02,LevinTwins01}.
This could be of
considerable importance to the standard cosmological model, since it
would provide a novel physical (geometrical) motivation for the existence of
a favoured space-time foliation, i.e. the comoving coordinate system.

%which otherwise is only an {\em ad hoc} postulate (Weyl's Postulate).

The difference between the twin paradox of special relativity
in a multiply connected space relative to that 
in a simply connected space is that in a multiply connected space,
the two twins can move with constant
relative speed and meet each other a second time,
{\em without requiring any acceleration}.
The paradox is the apparent symmetry
of the twins' situations despite the time dilation effect 
expected due to their non-zero relative speed.
It is difficult to understand how one twin can be younger than the other ---
why should moving to the left or to the right be somehow favoured?
Does the time dilation fail to occur?

As shown by several authors 
\citep{Peters83,Low90,UzanTwins02,LevinTwins01}, 
the apparent symmetry 
is violated by the fact that (at least) one twin must identify 
the faces of the fundamental domain of the spatial 3-manifold
{\em non-simultaneously}, and has problems in clock synchronisation.

\cite{UzanTwins02} suggested that the apparent symmetry is also
violated by an asymmetry 
between the homotopy classes of the worldlines of the two twins.

Here, we reexamine this suggestion.

The multiply connected space version of the twins paradox
is briefly reviewed in \SSS\ref{s-paradox}, 
the question of clarifying the asymmetry is described in 
\SSS\ref{s-asymmetry}, and 
a multiply connected space-time, with a standard Minkowski covering space-time,
and the corresponding space-time diagrams,
are introduced in \SSS\ref{s-st-diagrams}.
In \SSS\ref{s-worldlines-etc}, the worldlines of the two twins, 
projections from space-time to space,
homotopy classes and winding numbers 
are expressed algebraically.

In \SSS\ref{s-results}, the resulting 
projections of the twins' paths from space-time
to space are given and their homotopy classes inferred.
We also calculate whether or not the generator of the multiply connected space
and the Lorentz transformation commute with one another, in 
\SSS\ref{s-lorentz}.

%In \SSS\ref{s-discuss}, the results of this calculation are discussed,
Discussion suggesting why this result differs from that of 
\cite{UzanTwins02} is presented in \SSS\ref{s-discuss},
including a brief description in \SSS\ref{s-tperiodic}
of how the non-favoured twin
can choose a time-limited instead of a space-limited fundamental 
domain of space-time.
Conclusions are summarised in \SSS\ref{s-conclu}. A thought experiment
to develop physical intuition of the non-favoured twin's understanding 
of the spatial homotopy classes is given in 
Appendix~\ref{s-stretchable-cord}.

For a short, concise review of the terminology, geometry and 
relativistic context of cosmic topology (multiply connected universes in the
context of modern, physical cosmology), see \cite{Rouk00BASI} 
(slightly outdated, but sufficient for beginners).
There are several in-depth review papers available
\citep{LaLu95,Lum98,Stark98,LR99} and workshop proceedings 
are in \cite{Stark98} and the following articles,
and \cite{BR99}. 
Comparisons and classifications of different {\em observational} strategies,
have also been made \citep{ULL99b,LR99,Rouk02topclass,RG04}. 

We set the conversion factor between space and time units to unity
throughout: $c_{\mbox{space-time}}=1$.

\section{Method: space-time diagrams and worldlines}
\label{s-method}

The first articles presenting the special
relativity twins paradox in a multiply connected space were apparently
\citet{BransStewart73} and \citet{Peters83}. 
There are several more recent articles, including 
e.g. \citet{UzanTwins02} and \citet{LevinTwins01}.

\subsection{The twins paradox in multiply connected space} \label{s-paradox}

Consider the following description. 

In a one-dimensional, multiply connected, locally Lorentz invariant
space, one twin moves to the left and one to the right in rockets
moving at constant relative speed to one another. The two twins meet
twice, at two distinct space-time events.  At the earlier space-time
event, the two twins are of equal ages. At the later space-time event,
each twin considers the other to be younger due to Lorentz time
dilation.

However, this later space-time event is a single space-time event ---
each twin has undergone physical aging processes. If necessary, each twin
could carry an atomic
clock in order to more precisely measure proper time than with 
biological clocks.
So there can only be one ordinal relation between the two twins' ages at the
second space-time event: either the leftward moving twin is younger, 
or the rightward moving twin is younger, or the two twins are of equal 
age.\footnote{No quantum mechanical effects are considered in this paper.}
Which is correct?

There is no acceleration (change in velocity) by either twin, so
the usual explanation of the paradox (in simply connected space)
is invalid.

However, in this case, 
the situation is, or at least seems to be,
perfectly symmetrical. 
It would be absurd for either ``leftwards'' or ``rightwards'' 
movement to yield a younger age.

On the other hand, time dilation implies that the ``other'' twin must
``age more slowly''. The twins physically meet up (for an instant) at
the second space-time event, which is a single location in space-time,
so the ``first'' twin objectively measures that the ``other'' twin is younger,
so the two twins cannot be equally aged at the second space-time event.
But then which twin is ``the first'' and which is ``the other''?

This question illustrates the apparent 
symmetry of the situation and the apparent implication
that ``time dilation fails''.

Alternatively, if the situation is {\em not} symmetrical and time dilation
occurs as is expected, then what
breaks the apparent symmetry?  Why should leftwards movement by
favoured relative to rightward movement, or vice-versa?

Which is correct: is the situation symmetrical with a failure of time
dilation, or is the situation asymmetrical with time dilation taking place?

%%%\subsection{The solution: simultaneity is not absolute}

\subsection{Where is the asymmetry?} \label{s-asymmetry}

In 
\citet{Peters83},
\citet{Peters86},
\citet{UzanTwins02} and
\citet{LevinTwins01} it was 
shown that the apparent symmetry in the question as stated above is 
not mathematically (physically) possible. There is a hidden implicit 
assumption related to the usual intuitive error common to beginners in 
special relativity: the assumption of absolute simultaneity.

The necessary asymmetry can be described in different ways.
One way of explaining the asymmetry is as follows.

One twin is able to consistently synchronise her clocks 
by sending photons in opposite directions to each make a loop around the
Universe and observing their simultaneous arrival time,
and the other twin measures a delay between
receiving the two photons (or coded signal streams) 
and is forced to conclude that something is
asymmetrical about the nature of her ``inertial'' reference frame 
\citep{LevinTwins01}.

Carrying out this test requires a metrical measurement, that of proper time
intervals.

Here, in order to examine \cite{UzanTwins02}'s suggestion that there is 
a homotopy asymmetry, i.e. an asymmetry of topology rather than an asymmetry
depending on the metric,
it is easier to first explain the asymmetry of the apparently symmetrical
paradox in a more geometric way, similar to the 
presentation in \cite{Peters83}, but with some additional figures.

\fspacetimeone

\subsection{Multiply connected space-time diagrams} \label{s-st-diagrams}

Figure~\ref{f-spacetimeone} shows a standard Minkowski space --- as
covering space --- for simplicity with only one spatial dimension, 
for two twins moving with constant velocity 
relative to one another, hereafter,
the ``leftmoving'' and ``rightmoving'' twins respectively.
Specifically, we set the rightmoving twin moving at 
a velocity $\beta$, in units of the space-time
conversion constant $c \equiv 1$, towards the right. 
As a covering space, this space is a standard locally and globally
Lorentz invariant space-time $M$.

We choose a generator $g$ which favours, arbitrarily, but without loss of
generality, the leftmoving twin.
This generator, $g$, a translation of constant length  
$L$ with
\begin{equation}
g((x,t))= (x+L,t), \;\; \forall (x,t),
\label{e-generator}
\end{equation}
 generates the quotient, multiply connected space, 
$M/\Gamma$, where $\Gamma$ is the group generated by $g$, i.e. 
$\Gamma = \mathbb{Z}$.

This arbitrary choice reveals where an implicit assumption was made in 
the presentation of the paradox above: a generator matching space-time
events in a way that preserves time unchanged in one reference frame, or in 
other words, a generator which ``is simultaneous'' in one reference frame, 
is not simultaneous in other frames. Hence, symmetry is not possible.

\fspacetimetwo

The generator $g$ identifies 
points (in three-dimensional space, these would be 
faces of the fundamental domain rather than points)
in a spatial section at any given time $t$: 
A$_1=$A$_2$ and 
B$_1=$B$_2=$B$_3$. 

The rightmoving twin has $x'$ and $t'$ axes different from those
of her leftmoving twin, in order to preserve Lorentz
invariance. She disagrees with the leftmoving twin about simultaneity of events,
finding, e.g. that space-time event A$_2$ occurs before space-time event A$_1$,
space-time event B$_2$ occurs before space-time event B$_1$, etc. 
This is shown in Fig.~\ref{f-spacetimetwo},
where $\gamma \equiv (1-\beta^2)^{-1/2}$ is the Doppler boost.

So far, this is identical to the situation in 
simply connected Minkowski space-time, 
until we realise that both twins must agree that 
A$_1=$A$_2$ and that B$_1=$B$_2=$B$_3$. 

\fcylinderone

\fcylindertwo

While both twins agree that A$_1=$A$_2$, they {\em disagree}
as to whether or not these are simultaneous events. Using the terminology of
\cite{LevinTwins01}, the leftmoving twin is able to synchronise clocks,
while the rightmoving twin is {\em unable to synchronise clocks}.

Writing the Lorentz transformation from $(x,t)$ to $(x',t')$ as a matrix,
\begin{equation}
  {\cal L} = \left( 
  \begin{array}{cc}
    \gamma & -\beta \gamma \\
    -\beta \gamma  & \gamma\\
  \end{array}
  \right),
\label{e-lorentz}
\end{equation}
the generator $g$, initially expressed in the first twin's coordinates in 
Eq.~(\ref{e-generator}), can be rewritten using the second twin's coordinates
as
\begin{equation}
%g((x',t')) = \left(x' + \gamma L, t - \beta\gamma\frac{L}{c} \right),
g((x',t')) = \left(x' + \gamma L, t' - \beta\gamma L \right).  %% c=1 :)
\label{e-generator-moving}
\end{equation}

An intuitive, geometric way of describing this is that according 
to the rightmoving 
twin, {\em after
  cutting the covering space, non-simultaneous points are pasted together}.

Figure~\ref{f-cylinderone} shows the cylinder ``cut and pasted together'' 
out of a space-time region with constant space and time boundaries according
to the leftmoving twin. Note that a trapezium in Fig.~\ref{f-spacetimeone}, 
e.g. A$_1$A$_2$B$_3$B$_2$ 
would serve just as well as the rectangle  A$_1$A$_2$B$_2$B$_1$ for this 
``rolling up'' process. As long as there are boundaries of constant time $t$,
the result of identifying the other two sides of the trapezium is a cylinder.

This trapezium, A$_1$A$_2$B$_3$B$_2$, is particularly interesting when
we shift to the reference frame of the rightmoving twin in
Fig.~\ref{f-spacetimetwo}, since the boundaries A$_1$B$_2$ and
A$_2$B$_3$ now become spatial boundaries:
\begin{eqnarray}
t'(\mbox{A}_1)&=& t'(\mbox{B}_2) \nonumber \\
t'(\mbox{A}_2)&=& t'(\mbox{B}_3). 
\end{eqnarray}

Cutting and pasting from the rightmoving twin's point of view must still identify
identical space-time events to one another: {\em either} identifying 
 A$_1$B$_1$ to A$_2$B$_2$, or
 A$_1$B$_2$ to A$_2$B$_3$, will correctly apply the isometry to the covering
space and ``paste'' together our spatially finite interval in order to obtain
a manifold without any boundaries. (The time domain extends infinitely from this 
rectangle in both the negative and positive directions.)

So, one option for embedding this identification in 3-D Euclidean space and
projecting it onto the page would be to use the same trapezium.

This corresponds to the rightmoving twin's intuition of identifying
``two spatial points'' to one another while trying to ignore the nature of
space-time as a two-dimensional continuum: the
set of points along the line segment A$_1$B$_2$ constitute a single ``spatial
point'' $x'=0$, while the set of points along the line segment A$_2$B$_3$
constitute a single ``spatial point'' $x' = \gamma L$,
% where $\gamma
%\equiv (1-\beta)^{-1/2}$ is the usual Doppler boost. In space-time thinking,
a ``spatial point'' is really a worldline --- it is not just
a single point, it's a curve in space-time. 

However, rather than identifying the ``spatial borders'' of A$_1$A$_2$B$_3$B$_2$
to one another, it is helpful to follow the rightmoving twin's 
%na\"{\i}ve
naive intuition even further. 

Let us try to cut out and then paste together 
the space-time region with both constant space 
boundaries and constant time boundaries, i.e. the region A$_1$CDB$_2$.

The result is shown in Fig.~\ref{f-cylindertwo} 
(cf. fig.~5b in \citealt{Wucknitz04}).

This clearly shows the non-simultaneity of the cutting/pasting process for 
the rightmoving twin. The rectangle in $(x',t')$ space-time has to be given 
a time mismatch when it's pasted together.

This visually illustrates the error in 
the twins paradox in a multiply connected space (as described
in \SSS\ref{s-paradox}): 
{\em the implicit assumption of absolute simultaneity}.
If we implicitly assume absolute simultaneity, then we implicitly
assume that there is no inconsistency in supposing that both twins can
identify spatial boundaries without any time offsets. However, Lorentz
invariance is inconsistent with absolute simultaneity; hence, the asymmetry: at
most one twin can simultaneously identify spatial boundaries. Of course, 
neither the leftmoving twin nor the rightmoving twin are necessarily
favoured. A complete, precise statement of the problem needs to 
arbitrarily favour
one twin over the other: either the leftmoving or 
the rightmoving twin may be chosen, but one of them must be chosen to be
favoured in order for the space-time to be self-consistent.

This also illustrates the potential interest for cosmology: several authors 
note the existence of a favoured inertial reference frame implied by 
the multiple-connectedness of a (static) space-time whose covering space-time
is Minkowski and that this could provide a physical motivation for 
the comoving coordinate system \citep{UzanTwins02,LevinTwins01}.

\subsection{Worldlines, projections, homotopy classes and winding numbers}
\label{s-worldlines-etc}

Is the apparent (erroneous) symmetry of the two twins' situations
broken by asymmetry between the homotopy classes of the two twins'
projected worldlines (``spatial paths'') in some way?
In \cite{UzanTwins02} it was suggested that the twin who simultaneously
identifies spatial boundaries (in this paper, the leftmoving twin) has a
spatial path of zero winding index, while the rightmoving twin has a
spatial path of non-zero winding index.

\subsubsection{Worldlines}
\label{s-worldlines}

We consider the worldlines of the two twins between the two space-time events 
A and B, using Figs~\ref{f-spacetimeone} and \ref{f-spacetimetwo} 
and Eqs~(\ref{e-generator}), (\ref{e-lorentz})  and 
(\ref{e-generator-moving}). 
A superscript ``L'' or ``R'' is used to denote which reference frame
is being used for expression in a particular coordinate system.

The leftmoving twin's worldline,
$p = \overline{\mbox{A$_1$B$_1$}},$ 
can be written in her own reference frame as
\begin{eqnarray}
p^L &= & \left\{ ( 0, \tau) : 0 \le \tau \le \frac{L}{\beta} \right\}, 
\label{e-p-L}
\end{eqnarray}
using her proper time $\tau$ 
(see Fig.~\ref{f-spacetimeone}). 

A copy of the leftmoving twin's worldline as mapped by $g$ can be written, 
using Eq.~(\ref{e-generator}), as
\begin{eqnarray}
  p_g^L &\equiv& \{ g((x,t)) : (x,t) \in p \} \nonumber \\
  &        = & \left\{ ( L, \tau) : 0 \le \tau \le \frac{L}{\beta} \right\}, 
\label{e-pg-L}
\end{eqnarray}
We can write this in the rightmoving twin's frame using 
Eq.~(\ref{e-lorentz}):
\begin{eqnarray}
  p_g^R &= & \left\{ \gamma (L - \beta \tau, -\beta L  + \tau) : 
    0 \le \tau \le \frac{L}{\beta} \right\} .
\label{e-pg-R}
\end{eqnarray}
%and converting, for convenience, to proper time $\tau'$ in the 
%rightmoving twin's frame to be consistent with $\tau'$ in 
%Eq.~(\ref{e-q-R}) below, i.e. using $\tau' = \gamma\tau - \beta \gamma L$,
%we get
%\begin{eqnarray}
%  p_g &= & \left\{ \left( \beta \tau' + \frac{L}{\gamma}, 
%                                        \frac{\tau'}{\gamma} \right) : 
%    -\beta \gamma L \le \tau' \le \frac{L}{\beta \gamma} \right\} 
%\label{e-pg-R-old}
%\end{eqnarray}

The rightmoving twin's worldline,
$q = \overline{\mbox{A$_1$B$_2$}},$ 
can be written in her own reference frame as
\begin{eqnarray}
q^R &= & \left\{ (0, \tau') : 0 \le \tau' \le \frac{L}{\beta \gamma} \right\} 
\label{e-q-R}
\end{eqnarray}
using her own proper time $\tau'$ 
(see Fig.~\ref{f-spacetimetwo}), or equivalently in the leftmoving twin's frame,
using ${\cal L}^{-1}$ and Eq.~(\ref{e-lorentz}), 
as
\begin{eqnarray}
q^L &= & \left\{ (\beta \gamma \tau', \gamma \tau') : 
         0 \le \tau' \le \frac{L}{\beta \gamma} \right\} .
\label{e-q-L}
\end{eqnarray}

\subsubsection{Projections from space-time to space}

\cite{UzanTwins02} state that ``the asymmetry between [the two twins']
spacetime trajectories lies in a topological invariant of their
spatial geodesics, namely the homotopy class''.

In order to consider spatial geodesics in the two twins' reference frames,
we need to project from space-time to space in each of their reference 
frames.

Let us write these projections $\phi$ and $\phi'$ algebraically 
in the leftmoving and rightmoving twins' frames in the most obvious way:
\begin{eqnarray}
\phi((x,t))  &=& x
\label{e-phi-L}
\end{eqnarray}
and
\begin{eqnarray}
\phi'((x',t'))  &=& x'.
\label{e-phi-R}
\end{eqnarray}

Using Eqs~(\ref{e-generator}) and(\ref{e-generator-moving}),
the generator $g$ can then be rewritten in {\em projected} form as
\begin{equation}
g_\phi(x)= x+L, \;\; \forall x \in \mathbb{R},
\label{e-generator-proj}
\end{equation}
and
\begin{equation}
%g((x',t')) = \left(x' + \gamma L, t - \beta\gamma\frac{L}{c} \right),
g_{\phi'}(x') = x' + \gamma L, \;\; \forall x' \in \mathbb{R},
  %% c=1 :)
\label{e-generator-moving-proj}
\end{equation}
in the two twins' frames respectively.

The two spaces which space-time is projected to 
are clearly both equivalent to $\mathbb{R}/\Gamma \equiv S^1$, where the group 
$\Gamma$ is defined by the generator $g_\phi$ or $g_{\phi'}$ for 
the leftmoving or rightmoving twin respectively.
The length (circumference) of $S^1$ is 
$L$ or $\gamma L$ respectively.

It is important to note here that 
{\em in the absence of metric information, the two twins do not 
know anything more than this about the nature of their respective spaces,
and they do not know how their spaces relate to spatial
sections, i.e. subspaces of the covering space $M$.} 

In fact, metric information gives an important difference between
these two (in this paper, one-dimensional) spaces: the rightmoving
twin's constant time hypersurface has a time mismatch after ``going
around'' the universe once. However, here the question is to examine the
{\em topological} nature of the twins' spatial paths in the {\em absence} of metric
information. The space-time point of view which the rightmoving twin
would have, if she had metric information available (e.g. by sending
out photons and measuring proper time intervals), is irrelevant.
%discussed 
%in \SSS\ref{s-results-two-hypersurface}, and as a thought experiment
%in Appendix~\ref{s-stretchable-cord}. 

%In \SSS~III of \citet{UzanTwins02}, 

\subsubsection{Homotopy classes and winding numbers}
\label{s-winding-defn}

Similarly to \citet{UzanTwins02}, we write that two loops in 
$\mathbb{R}/\Gamma $ at A are homotopic 
to one another, 
\begin{equation}
p_1 \sim p_2 
\label{e-defn-homotopy}
\end{equation}
``if they can be {\em continuously} deformed 
into one another, i.e.  that there exists a continuous map'' 
$F : [0,1) \times [0,1) \rightarrow \mathbb{R}/\Gamma $ satisfying
\begin{eqnarray}
\forall \tau \in [0,1) &,& F(\tau,0) = p_1(\tau), F(\tau,1) = p_2(\tau) 
\nonumber \\
\forall s \in [0,1) &,& F(0,s) = F(1,s) = A.
\label{e-homotopymap}
\end{eqnarray}

As in \citet{UzanTwins02}, we will assume that no proof is needed
that $S^1 \not \sim \{ 0 \}$, 
and we define the winding number as the ``the number of times a loop rolls
around'' the circle $S^1$. Since the aim is to see if the leftmoving
and rightmoving twins' points of view can be distinguished in order to
determine who is favoured, we need
indices on the winding number to indicate whether it refers to the
worldline of the 
leftmoving or rightmoving twin, and from whose point of view. 

Again, we use the superscript ``L'' or ``R'' to denote which reference frame
is being used. A twin's point of view of her
own winding number is denoted by an empty subscript, 
and the subscript ``other'' is used for a twin's point of view of the
other twin's winding number. That is, the leftmoving twin measures her
own winding number $N^L$ and the other twin's winding number $\Nother^L$,
and the rightmoving twin measures her own
winding number $N^R$ and the other twin's winding number $\Nother^R$.

In each case, the winding number is positive in the direction of motion
of the ``other'' twin.

\fcylinderonehomotopy

\fcylindertwohomotopy

\section{Results}
\label{s-results}

\subsection{The leftmoving twin}
\label{s-results-one}

Figure~\ref{f-cylinderone_homotopy} shows that 
$p \equiv \overline{\mbox{A$_1$B$_1$}}$ projects to a point, and 
$q \equiv \overline{\mbox{A$_1$B$_2$}}$ projects to a closed loop. As stated in
\cite{UzanTwins02}, the former path has a zero winding index, while the
second has a unity winding index: from the point of view of the leftmoving
twin, there is a difference in the two twins' winding numbers. However, 
we haven't yet checked if the winding numbers are relative or absolute.

We evaluate the winding numbers as follows. Using the two twins' 
worldlines from the leftmoving twin's point of view, Eqs~(\ref{e-p-L}) and
(\ref{e-q-L}), and projecting them with Eq.~(\ref{e-phi-L}), we have 
\begin{equation}
\phi(p) = \phi(p^L) = \{ 0 \}
\end{equation}
and 
\begin{eqnarray}
  \phi(q) = \phi(q^L) &=& 
  \left\{ \beta\gamma\tau' : 0 \le \tau' \le \frac{L}{\beta\gamma} \right\} 
  \nonumber \\
          &=& \{ \lambda :  0 \le \lambda \le L \}.
\end{eqnarray}

Since $g_\phi(0) = L$ from Eq.~(\ref{e-generator-proj}), clearly 
\begin{equation}
\phi(p) \sim \{ 0 \}
\end{equation}
and
\begin{equation}
\phi(q) \sim S^1
\end{equation}
in the sense of Eq.~(\ref{e-defn-homotopy}),
so that the winding index of the leftmoving twin is zero, i.e.
\begin{equation}
N^L = 0
\label{e-NL}
\end{equation}
and that of the
rightmoving twin is unity,
\begin{equation}
\Nother^L = 1 .
\label{e-NL-other}
\end{equation}

\subsection{The rightmoving twin}
\label{s-results-two}

Figure~\ref{f-cylindertwo_homotopy} shows that 
$q \equiv \overline{\mbox{A$_1$B$_2$}}$ projects to a point, and 
$p_g \equiv \overline{\mbox{A$_2$B$_2$}}$ projects to a closed loop. 
The former path has a zero winding index, while the
latter has a unity winding index: from the point of view of the rightmoving
twin, 
there is a difference in the two twins' winding numbers. However, 
if each twin each assume that she is stationary and that the other is 
moving away, as we have assumed throughout, then they also 
disagree regarding the winding numbers: neither simultaneity
nor winding number is absolute.

We evaluate this algebraically as follows. Using the two twins' 
worldlines from the rightmoving twin's point of view, Eqs~(\ref{e-pg-R}) and
(\ref{e-q-R}) and projecting them with Eq.~(\ref{e-phi-R}), we have
\begin{eqnarray}
  \phi'(p_g) = \phi'(p_g^R) &=&
  \left\{ \gamma (L - \beta \tau) : 
  0 \le \tau \le \frac{L}{\beta} \right\} 
  \nonumber \\
  &=& [ \gamma L, 0 ] 
\label{e-pg-R-projected}
%  \left\{ \beta \tau' + \frac{L}{\gamma} : 
%    -\beta \gamma L \le \tau' \le \frac{L}{\beta \gamma} \right\} 
\end{eqnarray}
and 
\begin{eqnarray}
  \phi'(q) = \phi'(q^R) &=&  \left\{  0 \right\}.
\label{e-q-projected}
\end{eqnarray}

Eq.~(\ref{e-q-projected}) shows 
that in this case, the rightmoving twin has winding index zero:
\begin{equation}
N^R = 0 .
\label{e-NR}
\end{equation}

%For the leftmoving twin, substitute and simplify 
%the start and end points of $\phi'(p_g)$:
%\begin{eqnarray}
%  -\beta (-\beta \gamma L) +\frac{L}{\gamma} &=& 
%    L ( \beta^2 \gamma + \gamma^{-1} ) \nonumber \\
%    &=&  \gamma L
%\end{eqnarray}
%and
%\begin{eqnarray}
%  -\beta \left(\frac{L}{\beta\gamma} \right) +\frac{L}{\gamma} &=& 0.
%\end{eqnarray}

Since $g_\phi'(0) = \gamma L$ from Eq.~(\ref{e-generator-moving-proj}),
Eq.~(\ref{e-pg-R-projected}) clearly gives
\begin{equation}
\phi'(p_g) \sim S^1.
\end{equation}
Using the sign convention which we use globally here for the $x$ or
$x'$ coordinates,
the winding index of the leftmoving twin in this case would be $-1$.
More usefully, 
using the winding number convention described in \SSS\ref{s-winding-defn},
i.e. the winding number is positive in the direction of motion of the
``other'' twin, this becomes
\begin{equation}
\Nother^R = +1.
\label{e-NR-other}
\end{equation}
%since the leftmoving twin winds one loop around the space in the same
%direction in which she is moving.

\subsection{Summary: both twins} \label{s-results-both}

In summary, from Eqs.~(\ref{e-NL}), (\ref{e-NL-other}), 
(\ref{e-NR}) and (\ref{e-NR-other}), 
we have
\begin{eqnarray}
N^L = 0 & \; & \Nother^L = 1 \nonumber \\
N^R = 0 & \; & \Nother^R = 1.
\label{e-winding-summary}
\end{eqnarray}

Clearly, the homotopy classes of the twins' worldlines do not reveal
the asymmetry of the system: each twin considers herself
to have winding index $N=0$ and the other twin to have $\Nother = 1$.

\subsection{Commutativity of the generator $g$ 
and the Lorentz transformation $\cal L$} \label{s-lorentz}

\citet{UzanTwins02} also suggested in their eqs~(6)--(8) 
that the generator $g$ does not commute with the Lorentz transformation.
Here we calculate this explicitly.

%\subsubsection{ $  g \circ {\cal L} $}

Equation~(\ref{e-lorentz}) gives
\begin{eqnarray}
  g \circ {\cal L} ((x,t)) &=& g \circ (\gamma x -\beta\gamma t,
                               -\beta\gamma x + \gamma t).
\label{e-gofL-one}     
\end{eqnarray}
%%
%% g((x',t')) = \left(x' + \gamma L, t' - \beta\gamma L \right).  %% c=1 :)
Applying the generator in the appropriate reference frame,
Eq.~(\ref{e-generator-moving}), gives
\begin{eqnarray}
%g \circ (\gamma x -\beta\gamma t,
%                               -\beta\gamma x + \gamma t)
%\nonumber \\
 g \circ {\cal L} ((x,t)) 
 &=& [ (\gamma x -\beta\gamma t) + \gamma L,
      (-\beta\gamma x + \gamma t ) - \beta\gamma L ] \nonumber \\
 &=& \gamma [ x -\beta t + L , t -\beta (x + L ) ] .
\label{e-gofL-two}     
\end{eqnarray}
%
%\subsubsection{ $  {\cal L} \circ g$}
%
Equation~(\ref{e-generator}) gives
\begin{eqnarray}
  {\cal L} \circ g ((x,t)) &=& {\cal L} \circ ( x+L, t).
\label{e-Lofg-one}
\end{eqnarray}
The Lorentz transformation, Eq.~(\ref{e-lorentz}), gives
\begin{eqnarray}
  {\cal L} \circ ( x+L, t) &=&
  [ \gamma (x + L) - \beta \gamma t ,
    -\beta\gamma (x+L) + \gamma t ] \nonumber \\
  &=&
  \gamma [ x -\beta t + L , t - \beta (x+L) ] \nonumber \\
&=& g \circ {\cal L} ((x,t)),
%\label{e-Lofg-two}
\label{e-gofL-is-Lofg}
\end{eqnarray}
where the last equality uses Eq.~(\ref{e-gofL-two}). These last two
equations give
\begin{eqnarray}
{\cal L} \circ g((x,t)) &=& g \circ {\cal L} ((x,t)).
\label{e-g-L-commutativity}
\end{eqnarray}
%Clearly, we have 
%\begin{eqnarray}
% {\cal L} \circ g ((x,t)), \;\; \forall (x,t) \in M
%\end{eqnarray}

This result is in contrast to \citet{UzanTwins02}'s suggestion that 
$g \circ {\cal L} \not= {\cal L} \circ g$.
It is not obvious why our results differ.
%Possibly, \citet{UzanTwins02} wrote $g$ as an algebraically identical
%function in both reference frames, rather than as 
%a coordinate-independent isometry of the Minkowski space-time
%covering space $M$. 
Here, we consider $g \in \Gamma$  to be a holonomy transformation
on the covering Minkowski space $M$, 
independently of any particular coordinate representation.
As in \citet{UzanTwins02}, we consider
${\cal L}$ to be a transformation ``from one frame of space-time
coordinates to another system''.

\section{Discussion} \label{s-discuss}

The results summarised in \SSS\ref{s-results-both}
%\SSS\ref{s-results-one} and \ref{s-results-two}
differ from the conclusion in \cite{UzanTwins02}, where
it was pointed out that since winding indices are topological
invariants, ``neither change of coordinates or reference frame (which
ought to be continuous) can change [the winding indices'] values'',
i.e.  the rightmoving twin and the leftmoving twin should agree that
the leftmoving twin has a zero winding index and the rightmoving twin
a unity winding index. 
%(Hereafter, we return to the original left-right convention.)

\fcylinderoneeasy

This argument does not take into account the nature of the projection from
space-time to space. The argument is correct in that the topologically
invariant nature of winding indices is valid in {\em space-time}, but
is not necessarily valid in the worldlines 
{\em after projection from space-time to space}. 
The projection from an $n$-dimensional manifold to an $(n-1)$-dimensional
manifold does not (in general) conserve all topological properties of subspaces:
while a continuous projection 
conserves continuity, it does not conserve non-continuity.

For example, consider a discontinuous 
path $p_1 = \{ (x,f(x)) \;\; \forall x \in [0,L) \}$ in the leftmoving twin's 
reference frame, where $f(x)$ is a step function, i.e. a function discontinuous
for some values $x \in [0,L)$. This projects under $\phi$
to a continuous, closed loop. Similarly in higher dimensions:
consider  $S^1 \subset \mathbb{R}^3$, which can be continuously deformed
into something resembling a mess of string with the two
ends tied together and not touching itself anywhere. 
Project this from Euclidean 3-space into the Euclidean 2-plane.
The projection will (in general) be a graph with many nodes,
not $S^1$.

In fact, the worldlines (e.g. those labelled 1, 2, 3, 4 in fig.~2 in
\citealt{UzanTwins02}) are all open curves in space-time, i.e. each of them
is in the same homotopy class as a single point.
Figures~\ref{f-cylinderone_homotopy} and \ref{f-cylindertwo_homotopy} show
that these open curves only become either a loop or a point after projection.
The projection can convert a discontinuity into a continuity 
(or a continuum of points into a single point).

In space-time, it is
the union of the two twins' worldlines (two different paths in space-time 
from space-time event A to space-time event B) 
which forms a closed curve, not either
worldline alone.
This is shown in Fig.~\ref{f-cylinderone_easy}.

This is why a ``change of coordinates or reference frame'' using 
a continuous function {\em can} change the winding index values of
two worldlines, if those two worldlines are projections from $(n+1)$-space-time
to $n$-space, considered in two different reference frames.

Nevertheless, with the use of metrical information and space-time rather
than just space, it is interesting to note some different possibilities
of what ``space'' could correspond to if thought of as a subspace of 
the covering space-time $M$.

\subsection{The nature of spatial sections differs for the two twins}
\label{s-results-two-hypersurface}

Let us now use our knowledge of metrical information for considering
the nature of ``space'', i.e. spacelike sections of space-time, for
the two twins.  

\fspacetimetwoconst

A spacelike section of space-time at constant time for the leftmoving
twin connects with itself and is clearly $S^1$.

\fcylindertwotperiodic

In contrast, 
a spacelike section of space-time at constant time for the rightmoving
twin does not connect with itself: there is a time offset
of $\beta\gamma L$ after one loop. 
For a spacelike section to connect with itself,
it would need to be at non-constant time. This is not a problem for
topological measurements in the absence of metrical information: 
the rightmoving observer who sends out slow moving probes equipped with neither
clocks nor rods and having speeds which cannot be precisely calibrated
will assume that space is $S^1$.

In the covering space, this could be most simply interpreted as replacing
Eq.~(\ref{e-phi-R}) by a projection from $M$ to a subspace $S^1 \subset M$
defined:
\begin{eqnarray}
\phi_{\mbox{\small sub}}'((x',t'))  &=& (x', -\beta x')
\label{e-phi-R-subspace}
\end{eqnarray}
in the rightmoving twin's reference frame, or equivalently,
\begin{eqnarray}
\phi_{\mbox{\small sub}}'((x,t))  &=& (x - \beta t, 0 )
\label{e-phi-R-subspace-L}
\end{eqnarray}
in the leftmoving twin's reference frame. In other words, the subspace 
$S^1$ of $M$ correspond to the rightmoving twin's interpretation of space
from topological measurements alone would most easily be modelled as 
a spatial hypersurface of constant time for the leftmoving twin.

\subsection{The fundamental domain may be ``periodic'' in time instead of space}
\label{s-tperiodic}

While the rightmoving (non-favoured) 
twin making only topological measurements does not
{\em know} that 
her ``space'' $S^1$ consists of a set of space-time points at differing
times $t'$, the nature of what would be her constant time hypersurface,
if she were able to determine it, 
reveals an interesting property of a
multiply connected space-time for a non-favoured observer: {\em the 
fundamental domain can be chosen by matching time boundaries instead of space 
boundaries.}
Figures~\ref{f-spacetimetwoconst} and
\ref{f-cylindertwo_tperiodic} illustrate this situation.

The fundamental domain of this space-time 
can either be described as the region
\begin{eqnarray}
D_{x'} &\equiv & [0,\gamma L) \times \mathbb{R},
\label{e-fd-xp}
\end{eqnarray}
expressed as the direct product of a spatial interval as the
first parameter and a time interval as the second,
or as 
\begin{eqnarray}
D_{t'} &\equiv & \mathbb{R} \times [0,\beta\gamma {L}),
\label{e-fd-tp}
\end{eqnarray}
again with space first and time second.

The former, $D_{x'}$, is the ``obvious'' fundamental domain, satisfying
\begin{eqnarray}
\bigcup \left\{ g^i(D_{x'}), \forall i \in \mathbb{Z} \right\} &=& M 
\nonumber \\
g^i(D_{x'}) \cap g^j(D_{x'}) &=& \o , \; \forall i \not= j, i, j \in \mathbb{Z}
\end{eqnarray}
where the generator $g$ is easiest to use if expressed
in the rightmoving twin's reference frame, i.e. 
as in Eq.~(\ref{e-generator-moving}), and $g^2 = g \circ g$, 
$g^{-1} \circ g = g \circ g^{-1} = I$, etc.

The latter, $D_{t'}$, is the more surprising fundamental domain, 
satisfying
\begin{eqnarray}
\bigcup \left\{ g^i(D_{t'}), \forall i \in \mathbb{Z} \right\} &=& M 
\nonumber \\
g^i(D_{t'}) \cap g^j(D_{t'}) &=& \o , \; \forall i \not= j, i, j \in \mathbb{Z},
\end{eqnarray}
again where $g$ is best expressed using Eq.~(\ref{e-generator-moving}).

What is surprising is that a verbal description of $D_{t'}$ could be
``time is periodic for the non-favoured twin''. However, this description
omits important information; a more complete description would be
``time is periodic, but with a spatial offset, for the non-favoured twin''. 

In other words, the ability to choose a time-limited fundamental domain 
does {\em not} mean that events periodically repeat themselves
from the rightmoving twin's point of view, since the generator of the 
periodicity does {\em not} yield an offset by a
vector $(0,\beta\gamma {L})$. What it yields is
a space-time offset by (an integer multiple of)
the vector $(\gamma L, -\beta\gamma L)$. 

We could also say that for a non-favoured twin, the space-time
periodicity is ``diagonal'' to the space-time axes, i.e.
%the generator $g$ considered as a translation vector in the frame of a
%non-favoured observer is not parallel to either of the coordinate axes.
\begin{eqnarray}
&  \forall a \not= 0, b\not=0, a,b \in \mathbb{R},  (x',t') \in M, 
  \nonumber \\
  & g((x',t')) \not= (x' + a,t'), \; g((x',t')) \not= (x', t' + b).
\label{e-diagonal}
\end{eqnarray}

\section{Conclusions} \label{s-conclu}
%conclusions

Finding an asymmetry in the 
twin paradox of special relativity in a multiply connected space 
is less obvious than in a simply connected space, since neither twin
accelerates. It was already known that the asymmetry required is 
the fact that (at least) one twin must identify space-time events
non-simultaneously and has problems in clock synchronisation.

Here, space-time diagrams and the corresponding  algebra
have been presented in order to examine 
whether or not the homotopy classes of the twins'
worldlines provide another asymmetry. They show that homotopy classes
(numbers of windings) do {\em not} show which of the two twins of the
twin paradox has a preferred status, contrary to what was previously
suggested: each twin finds her own spatial path to have zero winding
index and that of the other twin to have unity winding index (in the direction
of travel of the other twin).

Although the twins' apparent symmetry is broken by the need for the non-favoured
twin to non-simultaneously identify spatial domain boundaries, and by the
non-favoured twin's problems in clock synchronisation (provided that she
has precise clocks), the non-favoured
twin {\em cannot}  detect her disfavoured state by measuring the
topological properties of the two twins' worldlines in the absence of 
precise metric measurements with clocks or rods. 

On the other hand, a non-favoured twin capable of making precise metric 
measurements will notice many surprising properties of space-time. 
For example, a non-favoured twin could define the fundamental domain 
of space-time to be $\mathbb{R} \times [0,\beta\gamma {L})$ where
the first dimension is space and the second is time. In words, time would
be ``periodic with period $\beta\gamma L$ 
and a spatial offset when matching time boundaries'' 
as an alternative explanation to having space be 
``periodic with period $\gamma L$ and a time offset when matching 
space boundaries''.

%She will not 
%$g \circ {\cal L} \not= {\cal L} \circ g$ [Eq.~(\ref{e-gofL-is-Lofg})] 

%Generalising
%from the discussion in \cite{Peters83}, we note the property that 
%there exist pairs of distinct space-time events, for a non-favoured twin,
%which are {\em both} spacelike and timelike separated 
%in the covering space-time, i.e. the generator of her manifold relative
%to her covering space-time is {\em diagonal} with respect to her space-time axes. 

\section*{Acknowledgments}

%The authors thank an 
%anonymous referee for many very constructive
%comments.

We thank Agnieszka Szaniewska for a careful reading and useful comments
on this paper. St.B. acknowledges suport form Polish
Ministry of Science Grant No NN203 394234.

%Helpful discussions and comments from
%...
%were greatly appreciated. 

\subm{\clearpage}

%%%  LINES BELOW WILL BE REMOVED - only \end{document} is added afterwards
%%%%%%%%%%%%%%%%%%%%%%%%%%%%%%
%\nice{

% Relax: the two bibtex bibliography lines here get commented out if you use
% aa2ps, which itself calls nat2aa.pl
% Contact soft@adjani.astro.uni.torun.pl if you need help finding aa2ps
% and nat2aa.pl .
% 

%}    %%%%%%%%%%%%%%%%%%%%%%%%%%%%%%%%%%%%%%%%%%%  

\appendix

\section{Thought experiment: stretchable cord between the twins}
\label{s-stretchable-cord}

%\subsection{Thought experiment: stretchable cord between the twins}

Since we are interested in topology, suppose that neither twin has
{\em precise} measuring rods or clocks, though both twins may have
{\em very approximate} methods of measuring metric properties.  Neither twin is
aware that when she completes a loop of the Universe, she may detect a
time offset. However, both twins have read history books and are aware
of claims that ``space is multiply connected'', so they attempt to 
verify this experimentally.

We can imagine that at event A, the two twins instantaneously create a
highly stretchable physical link between them, such as a light-weight
string or cord of negligible mass and extremely high strength against
breaking. As they move apart, the cord stretches, preserving the
topological properties of space connectedness, while ignoring time.

We could alternatively imagine that one of the twins leaves behind a
``trail'' of some sort, e.g. like the vapour trail of an aircraft
visible by human eye from the ground. However, in this case, we have 
to be careful to avoid thinking of the particles in the vapour trail
as being at rest in any particular frame, since otherwise we favour one
of the two twins arbitrarily. 

Now consider the state of the cord ``at'' event B, when the two twins
meet up again and join the two ends of the cord together. 
B is a single space-time event. Even though the two twins
disagree about space-time coordinates of the event, they
agree that it is a single event and agree that they have physically joined
the two ends of the cord.
Clearly the cord now forms a closed loop, of winding
index one. 

Note that the word ``at'' is, in fact, misleading, for two
reasons. 

Firstly, it is misleading because event B is just one space-time point 
among a whole set of space-time points
where the particles constituting the cord are located, 
but ``the state of the cord'' is only of interest at this point of the discussion
in the local neighbourhood of event B. It is difficult to
avoid intuitively thinking of ``the state of the cord'', i.e. of the
state of a spatially extended object ``at'' the {\em time} of the
event B, which is wrong, because it assumes simultaneity. A better way of
thinking of the cord is presented below.

Secondly, it is misleading because 
the word ``at'' suggests a mono-valued time coordinate for
a single event.  The reality is that just as in a multiply connected
space, a single (physical) spatial point exists at many spatial points in the
covering space, the situation is similar in a multiply connected space-time: a 
twin (observer) finds that a single space-time event exists at many
(in general) non-simultaneous space-time points in the covering space-time.
One twin happens to be favoured and finds that the multiple space-time copies
of a single event are simultaneous, but the other twin, moving with a different 
velocity, has a generator which is ``diagonal'' to her space-time axes.

Figs~\ref{f-spacetimeone} and \ref{f-spacetimetwo} can help to 
understand the space-time nature of the cord and to avoid the implicit
assumption of simultaneity.

From the
leftmoving twin's point of view, 
in Fig.~\ref{f-spacetimeone},
the cord can always be considered as
a simultaneous object, i.e. a series of successive ``snapshots'' of
the cord consist of horizontal line segments joining
$\overline{\mbox{A$_1$B$_1$}}$ and $\overline{\mbox{A$_1$B$_2$}}$,
starting at A and sliding up to a final state of
$\overline{\mbox{B$_1$B$_2$}}$ which for the leftmoving twin, is the
state of the cord ``at'' the time of space-time event B.

The rightmoving twin's point of view is similar,
except that as can be seen in Fig.~\ref{f-spacetimetwo}, ``simultaneous''
snapshots of the cord, i.e. horizontal 
line segments joining
$\overline{\mbox{A$_1$B$_1$}}$ and $\overline{\mbox{A$_1$B$_2$}}$,
starting at A and sliding upwards, have a problem when the right-hand
end of the cord arrives at B$_2$. At this point, the left-hand end of the
cord has not yet arrived at B$_1$ --- according to the righmoving twin's
notion of simultaneity.

However, B$_1$ and B$_2$ are a single physical space-time event: the
space-time event
of joining the two ends of the cord together 
occurs {\em non-simultaneously} according to the rightmoving
twin. 

This is intuitively difficult to imagine.
One way that the rightmoving twin could think about this could
be that ``as'' the cord slides up from event A, it tilts in some way so that 
when/where the two ends of the cord are joined at B, the cord can be imagined
as stretched along the line segment $\overline{\mbox{B$_1$B$_2$}}$ --- along
a series of space-time events which are non-simultaneous. This requires the
use of some arbitrary affine parameter to define ``as'' for the rightmoving twin, 
i.e. a parameter that 
is something like time but is not physical time. Of course, the 
simplest option for this parameter is the leftmoving twin's time coordinate, 
but this does not make it any easier for the rightmoving twin to develop her
intuition about it.

If we consider the cord to be ``stretched'' rather than ``unrolled'', so that the
parts of the cord closest to each twin are (nearly) stationary with respect 
to that twin, and if the cord is created containing 
a mix of
isotopes of radioactive elements
in initially known proportions, then at event B, the proper times at the
two ends of the cord will be measurable by measuring the isotopal
mixes.

In this case, both twins will agree that not only the rightmoving twin has
aged less, but also that the end of the cord ``held'' by the rightmoving twin
is younger than the end of the cord ``held'' by the leftmoving twin. So 
although the joined-up cord forms a single closed loop, its non-simultaneous
nature is revealed by the discordant ages (isotope mixes) 
of the two ends that are joined up at B. 

This is dependent on the thought experimental setup requiring the cord to 
be locally (nearly) at rest with respect to each twin, i.e. the cord is 
``stretched''.
With a different experimental setup for the behaviour of the cord,
the aging of the cord occurs differently, and can be 
calculated by studying the worldlines of the particles composing
the cord.

Now that we have some way of seeing either twin's way of thinking of 
this closed loop from B$_1$ to B$_2$, 
whose path through
space (projection of worldline to a spacelike hypersurface) does this loop
represent? Each twin considers herself to be stationary, and the other
twin to be moving ``rightwards'' or ``leftwards'', respectively. So
each twin considers her own path through space to be a single point
--- a path of zero winding index --- and that of the other twin to be
a closed loop --- a path of winding index unity represented by the
cord. Each twin considers the {\em other} twin to have pulled and/or
stretched the cord so that eventually the two ends could be joined,
not herself.

This is just an intuitive way of thinking of the projections described
above: an observer in a spatially multiply connected, locally Lorentz 
space-time, who is unable to 
make high-precision spatial and temporal measurements but can measure
topological properties of space is unable to use the homotopy class 
of her spatial path (projected worldline) to detect the fact that she
is either a favoured or a non-favoured observer.

%Alternatively, the twins might place some sort of marker on the cord
%at or just an instant after space-time event A, such that this marker lies
%somewhere between the two twins. In this case, the two twins might be more 
%likely to realise and agree that they are connected by a single closed curve,
%but that

Of course, if we understand the full nature of this space-time, then
we can note that the nature of ``the cord'' for at least one of the 
twins is a cross-section through space-time at non-constant time, as noted
above.
This is necessary in order for event B to be a single event in space-time.
It can also to help to remember a key idea in resolving the ``pole in
the barn paradox'' of simply connected Minkowski space: neither a pole 
nor the door-to-door path of a barn is a one-dimensional object --- 
both are {\em two-dimensional space-time objects}. The ``length'' of any such
object depends on the choice of  the reference frame, or in other
words, the choice of spacelike cross-section. 

Yet another useful way of thinking of a pole is as a ``worldplane''
--- a collection of worldlines. Our ordinary intuition of a pole as a
one-dimensional object is due to our implicit assumption of absolute
simultaneity. We can think of the cord stretched between the two twins and
joined up at event B to be the entire filled-in area of the triangle 
${\mbox{A$_1$B$_1$B$_2$}}$ in Figs~\ref{f-spacetimeone} and 
\ref{f-spacetimetwo} --- a two-dimensional space-time object. 
Depending on various possible thought experimental
setups for creating/producing/stretching the cord, various sets of worldlines
for the particles composing the cord are possible, but in each case would 
fill in this triangle.

%\end{appendix}

%% If the twin who sees the joined cord as existing in a non-constant time
%% cross-section of space-time actually {\em knows} this, then this 
%% means that she has earlier sent out photons in both directions in order to 
%% calibrate her clocks.
%% 
%% In this case, she might also note that her hypersurfaces of constant 
%% time either need to be arbitrarily cut cover the whole 

\newcommand\unusedstuff{
\clearpage

\section{ALL OF THE FOLLOWING STUFF IS FROM OLDER VERSIONS. IT
WILL BE REMOVED UNLESS SOMEONE QUICKLY ASKS FOR IT TO BE PUT BACK IN THE
PAPER.}

{\em

A paradoxical aspect of Figure~\ref{f-spacetimetwoconst}, given 
normal relativistic
intuition, is that it shows that for the rightmoving 
twin, there exist certain
pairs of distinct events in the covering space-time for which
the two members of the pair
may be considered either as located in the same spatial position but separated
in time, or as simultaneous and separated in space, depending on which multiple
images of the events in space-time are chosen for the comparison 
\protect\citep[cf section III][]{Peters83}. 

Another way of describing this
is that a pair of events can be {\em both} spacelike and timelike separated in 
the covering space-time. This is due to the existence of multiple images in 
the covering space-time, and hence multiple separation vectors between a single
pair of physical events. 

If a particular choice of fundamental domain is made, with each event
occurring exactly once, then only one geodesic between the pair of events
exists entirely inside of the fundamental domain, and the ambiguity is removed. 
For example, for events A and C in $[0,\gamma L)
\times \mathbb{R}$ or $\mathbb{R} \times [0,\beta\gamma
\frac{L}{c})$, the separation vector is  
timelike or
spacelike respectively and there is no ambiguity. However, this is
an arbitrary choice. 

Although this property has partially been explored earlier in
\cite{Peters83}, it is useful to summarise it more generally as follows:

{\em For a non-favoured twin (a twin who identifies spatial fundamental domain 
boundaries non-simultaneously), there exist 
pairs of distinct events which are both spacelike and timelike separated
in the covering space-time.}

Let us return to the need to find a hypersurface on to which the rightmoving
twin's worldline can be projected.
If the rightmoving twin makes precise space-time measurements, using precise
clocks and rods, then she will notice that her constant time
hypersurfaces ``wrap around'' the whole of ``space'' many times, or,
in fact, infinitely many times if the local model is extended
globally (static space with an infinite time axis). 
This extends the discussion of the 
``pole in the Universe paradox'', a variant on the ``pole in the barn
paradox'' \citep[section V,][]{Peters83}. However, this type of hypersurface
can only be known to the twin if she has precise metric measuring instruments, 
and would not help her with topological measurements.

\subsubsection{Projection to a ``cross-sectional'' hypersurface}
\label{s-rightmovingproj}

If the twin is only interested in measuring {\em topological} properties of 
``space'', e.g. if she lacks
precise clocks for attempting clock synchronisation and lacks precise metre 
sticks for measuring distances, but can measure ``spatial'' topological
properties, 
then the relevant spacelike hypersurface onto 
which the worldlines can be projected should be one which is a spacelike 
``cross-section'' $X$ of the space-time 
$X\times \mathbb{R} = M/\Gamma$, for an infinite time domain, 
i.e. in the case illustrated in this discussion, $X = T^1 \equiv S^1$. 
%(These arguments clearly extend to $T^n$ where $n=2,3$.)

One obvious choice of such a cross-section is the hypersurface of constant 
time $t$ for the leftmoving twin, even though the rightmoving twin, lacking
precise measuring tools, may not be fully aware of the nature of this surface.

Figure~\ref{f-cylindertwo_homotopy} shows that when the rightmoving twin
projects into a spacelike hypersurface topologically equivalent to $X$, the
same spatial cross-section as in the leftmoving twin's point of view,
the situation is symmetrical to 
that of the leftmoving twin. In this projection, the rightmoving twin's
(projected) path has a winding index of zero, while the leftmoving twin
has a winding index of one (or minus one, if we include a sense of direction). 
%but hereafter we will ignore this).

\subsection{Projected worldlines of the two twins: which twin is favoured?}

To summarise: the projections of the worldlines of {\em either} twin
into a cross-sectional, spacelike hypersurface implies that the twin finds
herself to be following a path of winding index zero and considers the
``other'' twin to be following a path of non-zero winding index. 

More specifically, given the left-right convention used above in this
paper, we have the following.

The leftmoving twin considers herself to be stationary and to have a
winding index of zero, and considers the rightmoving twin to be moving to 
the right and to have a
winding index of $+1$.

The rightmoving twin considers herself to be stationary and to have a
winding index of zero, and considers the leftmoving twin to be moving to the
left and to have a winding
index of $-1$.

Is this difference in sign important? Does it reveal the asymmetry between
the two twins' situations? Does it show which twin is older?

So far in this paper, we have used an ``absolute'' convention on
labelling ``left'' and ``right'', in which ``left'' is the negative
(spatial) direction and ``right'' is the positive (spatial)
direction. To avoid confusion, we used the same convention for both
twins, so that one can be called ``leftmoving'' and the other
``rightmoving'', even though in reality, this is an arbitrary choice.
If we write the winding indices as $N_L$ and $N_R$ for the leftmoving
and rightmoving twins respectively, then we have $N_R-N_L = +1$,
independently of which twin makes the calculation. In both cases, the
rightmoving twin has a more positive winding index than the leftmoving
twin.

In some sense, this is an asymmetry, since one sign (positive) is favoured.
However, this is an artefact of our choice of sign convention. According 
to this choice of sign convention, 
one twin is relatively leftmoving and one is
relatively rightmoving. In other words, one is relatively
``negative-direction-moving'' and the other is relatively
``positive-direction-moving''. 

A more neutral convention would be to define ``right'' as the
``direction in which the other twin is moving''. This is a non-absolute
convention. 

With this convention, we can no longer distinguish the twins by
labelling one as leftmoving and one as rightmoving: each twin
considers herself to be stationary and the other twin to be moving
towards the right. All the above diagrams remain valid, except that
right and left need to be swapped in diagrams presented from the point
of view of the twin whom we earlier called ``rightmoving''. We can now
call the two twins who earlier were labelled as ``leftmoving'' and
``rightmoving''  as, respectively, the ``hitherto-leftmoving'' and
``hitherto-rightmoving'' twins.

For the hitherto-leftmoving twin, her winding index is still $N_L=0$
and the other twin's winding index is still $N_R=+1$
(\SSS\ref{s-results-one}, Figure~\ref{f-cylinderone_homotopy}).

The hitherto-rightmoving twin still has a zero winding index from her
own point of view (\SSS\ref{s-rightmovingproj},
Figure~\ref{f-cylindertwo_homotopy}), but since she considers herself
relatively leftmoving (according to the new sign convention), her
winding number is labelled $N_L$, so we have $N_L=0$. 

Moreover, she finds the winding index of the other twin to be
$N_R=+1$, since ``right'' is the direction of movement of ``the
other'' twin (Figure~\ref{f-cylindertwo_homotopy} is left-right reversed).

Thus, we have $N_L= 0, N_R=+1$ from the point of view of either twin,
so the artificial asymmetry introduced by the sign convention used earlier
in this paper disappears.

We remind the reader that the asymmetry we are searching for relates
to the problem of finding out which twin is older: a spatial direction
sign convention does not reveal this, since it is an arbitrary
choice. To quote \cite{UzanTwins02}, ``If space is compact, then a
traveller twin can leave Earth, travel back home without changing
direction and find her sedentary twin older than herself. We show that 
the asymmetry between their spacetime trajectories\ldots''

The paradox is to find an asymmetry which explains why one twin is
older than the other despite the fact that either twin can consider
herself to be the sedentary twin and the other twin to be the
traveller twin (the inclusion of the Earth is irrelevant). One example
of an objective, local measurement showing a difference between the
twins' situations is measuring their ages at event B. So the question
concerns what alternative measurement (or measurements), other than
measuring and comparing the twins' ages, can enable a twin to
determine that she is non-favoured (younger at event B).

What we have shown here is that the homotopy class of a twin's
worldline projected into a spacelike hypersurface does {\em not}
enable a twin to decide whether or not she is a favoured (older)
twin. It only enables her to decide that she has a projected worldline
with zero winding index and that the other twin has a projected
worldline with $+1$ winding index, where the $+$ indicates the
direction of travel of that other twin. Since both twins individually
find this same result, this is not a sufficient measurement to decide
which twin is favoured (older): topological properties of projected
worldlines are insufficient to decide which twin is favoured.

{\em POSSIBLY REUSE THE FOLLOWIGN SOMEWHERE......

Again, a single domain of the space-time, with constant spatial boundaries for
that observer, will be shown for each observer, as in 
Figs \ref{f-cylinderone} and \ref{f-cylindertwo}, but with the addition 
of the two worldlines and their projections into spacelike hypersurfaces.

%\subsubsection{Spacelike hypersurface on which to project wordlines}
%\label{s-projectionhypersurface}

The spacelike hypersurface onto which the worldlines will be projected for the
leftmoving twin is a constant time hypersurface, as in \cite{UzanTwins02}.

However, for the rightmoving twin, the choice of which spacelike
hypersurface should be used for the projection is ambiguous, and is presented
in more detail below in \SSS\ref{s-results-two}.
}

}  % stuff to grab or else throw out

} % \newcommand\unusedstuff{

%\unusedstuff{

\end{document}